\documentclass[fleqn,usenatbib]{mnras}

\usepackage[T1]{fontenc}

\DeclareRobustCommand{\VAN}[3]{#2}
\let\VANthebibliography\thebibliography
\def\thebibliography{\DeclareRobustCommand{\VAN}[3]{##3}\VANthebibliography}

\usepackage{graphicx}	
\usepackage{amsmath}	
\usepackage{amssymb}	
\usepackage{booktabs}
\usepackage{ulem}

\usepackage{lscape}

\title[NGC\,4214]{H$\alpha$ Kinematics of Superbubbles and Supernova Remnants  of the Dwarf galaxy NGC\,4214}

\author[M. S\'anchez-Cruces \& M. Rosado]{
M. S\'anchez-Cruces$^{1}$\thanks{E-mail: monica.sanchez.cruces@gmail.com}
and M. Rosado$^{1}$\thanks{E-mail: margarit@astro.unam.mx }
\\
$^{1}$Instituto de Astronom\'ia, Universidad Nacional Aut\'onoma de M\'exico, Circuito Exterior, C.U., Apartado Postal 70-264, 04510 \\
Ciudad de M\'exico, M\'exico\\
}

\date{Accepted XXX. Received YYY; in original form ZZZ}

\pubyear{2023}

\begin{document}
\label{firstpage}
\pagerange{\pageref{firstpage}--\pageref{lastpage}}
\maketitle
\begin{abstract}
We analysed the ionised gas kinematics of the dwarf galaxy NGC\,4214 using high resolution Fabry-Perot interferometry observations and present a set of narrowband images in the H$\alpha$, [\ion{S}{ii}] $\lambda$6717 \AA, [\ion{N}{ii}] $\lambda$6584 \AA\ and [\ion{O}{iii}] $\lambda$5007 \AA\ emission lines.
The high-resolution Fabry-Perot observations of the H$\alpha$ emission line, allowed us to derive the velocity field, the velocity dispersion $\sigma$, and the rotation curve of the galaxy. We also present for the first time, three-dimensional kinematic maps of the complexes NGC\,4214-I and NGC\,4214-II and analysed the kinematics of the ionised gas of two new superbubbles, as well as the supernova remnants previously detected in this galaxy by other authors, in radio, optical and X-ray emission. We computed the expansion velocities of the superbubbles and supernova remnants fitting their velocity profiles and obtained their respective physical parameters. 
We found that the superbubbles have an expansion velocity of $\sim$50 km s$^{-1}$, dynamical age about $\sim$2 Myr and wind luminosity L$_W$ of $\sim$9$\times$10$^{38}$ erg s$^{-1}$ produced probably by massive stars in OB associations.
For supernova remnants, their expansion velocities are between $\sim$48 to $\sim$80 km s$^{-1}$ with ages of about 10$^{4}$ years and kinetic energy of about 10$^{51}$ erg assuming they are in the {radiative} phase of evolution.

\end{abstract}




\section{Introduction}
\label{intro}

NGC\,4214 is a nearby \citep[2.7 Mpc,][derived from the galaxy distance from the tip of the red giant branch (TRGB)]{Drozdovsky2002, Lelli2014} irregular dwarf starburst galaxy also known as UGC 7278 and also classified as Magellanic type (NASA/IPAC Extragalactic Database (NED)).

NGC\,4214 has two main regions of star formation located in the central part of the galaxy; the northwest (NW) complex, also called NGC\,4214-I, is a large complex of H II regions displaying a shell morphology and the southeast (SE) complex known as NGC\,4214-II, formed mainly of compact knots  \citep[see for example][]{MacKenty2000,Fahrion2017,Choi2020}. Also, \cite{MacKenty2000} identified 11 more regions corresponding to nebular knots or stellar clusters with weak or intense nebular emission (named NGC 4214-III to NGC 4214-XIII). NGC\,4214-I has several star clusters, the most important is I-As, a massive young \citep[3.0-5 Myr][]{Leitherer1996, MacKenty2000, Ubeda2007} superstar cluster (SSC). NGC\,4214-II contains several Scaled OB Association (SOBAs)\footnote{Scaled OB Association (SOBAs) are quite asymmetric extended objects with no well-defined center with sizes larger than 10 pc \citep{MaizApellaniz2001}.}. Also, these complexes are known to be rich in Wolf-Rayet stars \citep{Sargent-Filippenko1991,MasHesse-Kunth1999, Drozdovsky2002,Srivastava2011} making it a Wolf-Rayet Galaxy. General parameters of the galaxy are shown in Table \ref{Table1}.\\

The study of the ionised gas of this galaxy can help understand the
interaction of their stellar clusters and massive stars with the Interstellar Medium (ISM). In particular, the formation mechanisms of objects at different scales such as bubbles, superbubbles and supershells. 

Bubbles are formed by the interaction of strong stellar winds of massive stars with its surrounding interstellar medium. In optical emission, 
these bubbles can be detected as ring shaped nebulae \citep[see ][]{Rosado1983, Goodrich1987, Marston1994} with diameters, D $\lesssim$ 10 pc.
Superbubbles (SBs) or giant shells are objects blown by the combination of fast stellar winds and supernova explosions from groups of massive stars \citep[see][]{Bruhweiler1980, TenorioTagle1988, MacLow1988, Chu2008} as ring-shaped structures with D $\lesssim$ 400 pc.
Supershells or supergiant shells are the result of the evolution of super star clusters (SSCs) and their interaction with the ISM of a galaxy with D $\approx$ 1 kpc \citep[see][]{Meaburn1980, Martin1997, Yamaguchi2001,RodriguezGonzalez2011, Dawson2013,ReyesIturbide2014}.

The kinematics of the ionised gas provides important clues to understand the hydrodynamic nature of the interrelationship between the ISM and those types of objects and, the physical processes that affect the surrounding ISM.
In the kinematical study performed by \citet{MaizApellaniz1999} on the ionised gas of the nuclear region of this galaxy, they found that 
the superbubbles associated to the stellar clusters in the central region are not according to their evolutionary state. They found that the most massive cluster (cluster A) presents a partial ring feature while cluster B has one shell that seems to have experienced blowout, and complete second shell. \cite{Wilcots2001} using WIYN Integral Field Unit (IFU) spectroscopy found evidence of high-velocity outflows (50-100 km s$^{-1}$) associated with diffuse ionised gas and found a correlation between high-velocity gas and \ion{H}{I} holes in this galaxy.

Supernova remnant (SNR) multiwavelength surveys (X-ray, radio and optical) have been been carried out for NGC\,4214, providing information about the physical properties of these objects. In particular, it has been found that some SNRs present emission in the X-ray, radio and optical wavelengths \cite[see][]{Leonidaki2013}. A brief summary of the population of SNRs in each wavelength band is given below.

By using radio observations of the  Very Large Array (VLA), \cite{Vukotic2005}  classified one radio source as SNR; four years later, \cite{Chomiuk2009} radio continuum (VLA data at 20, 6, and 3.6 cm) and optical data, found six more radio SNR candidates and three three objects denoted as SNR/\ion{H}{ii}.
 
Then, \cite{Leonidaki2010}, based on the X-ray colors or spectra of the SNRs from Chandra archival data, suggested 11 X-ray sources as SNRs or SNR candidates.  
Finally, \citet{Leonidaki2013} found 18 confirmed optical SNRs based on the spectroscopy emission-line flux criterion of [\ion{S}{ii}]/H$\alpha$ >0.4.

Considering all previously published SNR samples in radio, X-ray, and optical emission, and that some of them present emission in two or three wavelength bands; this galaxy harbors a total of 35 SNRs which until now, there is no kinematic information of them (see Section \ref{Param_SNRs}). In this work we studied the global kinematics of the galaxy and the kinematics at local scales such as around the SBs and SNRs in the dwarf galaxy NGC\,4214 by using scanning Fabry-Perot (FP) observations.
Also, in this study, we present, for the first time, a three-dimensional kinematic map of the complexes NGC\,4214-I and NGC\,4214-II.

The structure of the paper is as follows. Section 2 presents the scanning FP observations and data reductions. Section 3 presents the general morphology of the galaxy. In Section 4 we present the kinematic information derived from the FP observations, and in Section 5  we present the conclusions.

\begin{table}
	\centering
	\caption{General Properties of NGC\,4214}
	\label{Table1}
	\begin{tabular}{lcc} 
\hline
Parameter &
Value &
Reference\\	
		\hline
Other names	&	UGC 07278 			&	NED	\\
RA (J2000)	&	12:15:39.17		&	NED	\\
Dec (J2000)	&	+36:19:36.8		&	NED	\\
Vsys		&	291 km\,s$^{-1}$	&	NED	\\
Distance	&	2.7 Mpc			&	NED	\\
Type		&	IAB(s)m				&	NED	\\
 i$_\text{opt}$			&	26$\degr$			&	\citet{Lelli2014}	\\
 PA$_\text{opt}$			&	40$\degr$			&	\citet{Lelli2014}	\\
 i$_\text{kin}$			&	30$\degr$			&	\citet{Lelli2014},\\ &&\citet{Swaters2009}\\
 PA$_\text{kin}$			&	65$\degr$			&	\citet{Lelli2014}	\\
 			&	74$\degr$			&	\citet{Swaters2009}  	\\
M$_{*}$	 (M$_{\odot}$)	&$>28 \times 10^7$&	\citet{Lelli2014}	\\
			&1.5 $\times 10^{9}$  & \cite{Karachentsev2004}	\\
SFR	(M$_{\odot} yr^{-1}$)		& 130 $\times 10^{-3}$  & \cite{Lelli2014} \\
M$_\text{\ion{H}{I}}$ ($M_{\odot})$ & 4.1 $\times 10^{8}$&	\citet{Walter2008}	\\
12 + log (O/H) & 8.22 & \citet{Kobulnicky1996}  \\
L$_\text{x}$ (erg s$^{-1}$) 	& 3.4$\times$10$^{38}$	 	& \citet{Hartwell2004}\\
L$_\text{FIR}$ (erg s$^{-1}$)		&5.3$\times$10$^{42}$& \citet{MasHesse-Kunth1999}\\
L$_\text{radio}$	(Jy kpc$^{2}$)		&3.6$\times$10$^{7}$	& \citet{MasHesse-Kunth1999} \\
		\hline
	\end{tabular}
\end{table}

\section{Observations and data reduction}

\subsection{Direct imaging}
The observations were carried out in November 27, 1998 using the 2.1 m telescope of the Observatorio Astronómico Nacional of the Universidad Nacional Autónoma de México (OAN, UNAM) at San Pedro Mártir, B. C., México. The observations were made in the H$\alpha$ line, using the UNAM scanning FP interferometer, PUMA \citep{Rosado1995}. 

The image size is 512 $\times$ 512 px produced by binning a 1024$\times$1024 Tektronix2 CCD detector by a factor of two. The field of view (FoV) of PUMA is 10 arcmin and has a plate scale of 1.16 arcsec pixel$^{-1}$ (15.18 pc pixel$^{-1}$ at a distance of 2.7 Mpc). 

The direct images were obtained with the FP etalon out of the telescope's line-of-sight (PUMA direct-imaging mode) in H$\alpha$, [\ion{S}{ii}] $\lambda$6717 \AA, [\ion{N}{ii}] $\lambda$6584 \AA\ and [\ion{O}{iii}] $\lambda$5007 \AA\ emission lines redshifted to NGC\,4214. The exposure time of the  H$\alpha$ image was 60 s and 120 s for the other emission lines. We also obtained direct images of the spectrophometric standard star \textit{Feige92} \citep{Stone1977} with an exposure time of 60 s in each filter. Table \ref{Table2} shows the journal of the PUMA observations. 

\subsection{High resolution Fabry-Perot spectroscopy}
The PUMA scanning FP interferometer  mode  has a finesse of $\sim$24 leads to a sampling spectral resolution of 0.41 \AA\ (equivalent to a velocity resolution of 19.0\,km\,s$^{-1}$ at H$\alpha$) and a free spectral range of 19.8 \AA\ (equivalent to a velocity range of 908\,km\,s$^{-1}$ at H$\alpha$). 

Then, with the  FP interferometer PUMA, we scanned the  H$\alpha$ line through 48 channels getting a data cube of dimensions  512 $\times$ 512 $\times$ 48.  The integration time was 60 s per channel. For the calibration of the data cube, we used a Hydrogen lamp (6562.78 \AA\ wavelength calibration); this calibration data cube has the same dimensions as the object data cube. Observational and instrumental parameters are listed in Table \ref{Table2}.

\begin{table}
	\centering
	\caption{Observational and Instrumental Parameters}
	\label{Table2}
	\begin{tabular}{llllll} 
\hline	
Parameter &
\multicolumn{4}{c}{Value} \\	
\hline
Telescope	&	\multicolumn{4}{c}{2.1 m OAN-UNAM}	\\
Instrument	&	\multicolumn{4}{c}{PUMA}	\\
Detector	&	\multicolumn{4}{c}{Tektronix2 CCD}	\\
Detector size (pix$^{2}$) 	&	\multicolumn{4}{c}{1024$\times$1024}	\\
Field of view	&	\multicolumn{4}{c}{10$\arcmin$}	\\
Image scale	&	\multicolumn{4}{c}{1.16 arcsec/pix}	\\
Observation date	& \multicolumn{4}{c}{November, 1998}	\\
\hline	
\multicolumn{5}{c}{Direct images} \\
\hline	
Filter						&	H$\alpha$ &[\ion{S}{ii}] &  [\ion{N}{ii}]  & [\ion{O}{iii}] \\
Central wavelength (\AA)	& 6570 & 6720 & 6584 & 5007	\\
Bandwidth (\AA)				&	20 & 20  & 10 & 10 \\
Exposure time (s)	& 60 	& 120 	& 120	& 120	\\
\hline	
\multicolumn{5}{c}{Fabry-Perot} \\
\hline	
Interference filter	&	\multicolumn{4}{c}{6570/20}	\\
Scanning steps	&	\multicolumn{4}{c}{48}	\\
Exposure time per channel	&	\multicolumn{4}{c}{60 s}	\\
Calibration line	&	\multicolumn{4}{c}{H$\alpha$}	\\
Interference order at H$\alpha$	&	\multicolumn{4}{c}{330}	\\
Free spectral range at H$\alpha$	&	\multicolumn{4}{c}{19 \AA\ (908 km\,s$^{-1}$)}	\\
Spectral sampling at H$\alpha$	&	\multicolumn{4}{c}{0.42 \AA\ (19.23 km\,s$^{-1}$)}	\\
\hline
	\end{tabular}
\end{table}

\subsection{Data reduction}

The reduction and analysis of FP data were performed
using IRAF\footnote{`Image Reduction and Analysis Facility' \url{http://iraf.noao.edu/}. IRAF is distributed by National Optical Astronomy Observatory, operated by the Association of Universities for Research in Astronomic, Inc., under cooperative agreement with the National Science Foundation.} and the ADHOCw\footnote{\url{http://cesam.lam.fr/fabryperot/index/softwares} developed by J. Boulesteix.} software. We made the standard corrections: i.e., removal of cosmic rays and bias subtraction. We computed the wavelength-calibrated data cube and used it to get the velocity\footnote{All quoted velocities are in the heliocentric reference frame.} and velocity dispersion maps. 

ADHOCw uses the calibration data cube to produce the phase map; this map provides for each pixel, the reference wavelength in the observed line profile. The wavelength-calibrated data cube (velocity or wavelength data cube), is produced when the phase map is applied to the observed data cube.

In order to improve the signal-to-noise (S/N) ratio, we applied on the wavelength data cube a spectral Gaussian smoothing using a FWHM of 3 channels ($\sigma$=57~km~s$^{-1}$), and spatial Gaussian smoothing using a FWHM of (3,3) pixels by using ADHOCw software.

The ``monochromatic'' and ``continuum'' images were obtained with ADHOCw by integrating the radial velocity profile from the FP data cube pixel by pixel in the FoV and the continuum image is derived pixel by pixel as the mean of the three lowest intensity channel, which typically corresponds to 30\% of the peak intensity value. For the monochromatic image, ADHOCw integrates the profile of the line in each pixel from its maximum intensity (I$_{max}$) up to 0.3$\times$I$_{max}$.

Finally, for each pixel ADHOCw computes the H$\alpha$ profile barycenter to obtain the velocity and velocity dispersion ($\sigma_\text{obs}$) maps.

We performed thermal and instrumental corrections to the velocity dispersion, according to: : $\sigma_\text{corr}$ = $\sqrt{\sigma_\text{obs}^{2} - \sigma_\text{inst}^{2} - \sigma_\text{th}^{2}}$ where
$\sigma_\text{inst}$=14.85~km~s$^{-1}$  is the instrumental broadening; $\sigma_\text{th}$=$\sqrt{82.5(\text{ T}_4/\text{A})}$ is the thermal broadening with T$_4$ = T/10$^{4}$~K and A the atomic weight of the atom, taking T=10$^{4}$~K for hydrogen gas \citep{Spitzer1978} we have $\sigma_\text{th}$ = 9.1 km s$^{-1}$.

The velocity accuracy for our data is about $\pm$2~km~s$^{-1}$ obtained from the calibration velocity map. This map was computed with the barycenter of the emission line in each pixel of the calibration data cube. 

\subsection{Photometric Calibration}
For the flux calibration of our direct images, we used the images of the standard star \textit{Feige92} taken in each filter with PUMA in its direct image mode. All direct images of the galaxy and standard star
were reduced as described above. 
The conversion factor for each emission line is  1 count s$^{-1}$ pixel$^{-1}$ = (2.92(H$\alpha$), 3.27([\ion{S}{ii}]), 5.90([\ion{N}{ii}]),  4.47([\ion{O}{iii}])) $\times$ 10$^{-4}$ erg cm$^{-2}$ s$^{-1}$ sr$^{-1}$, which in units of flux is (1.16(H$\alpha$), 1.30([\ion{S}{ii}]), 2.34([\ion{N}{ii}]), 1.77([\ion{O}{iii}])) $\times$ 10$^{-14}$ ergs cm$^{-2}$ s$^{-1}$ sr$^{-1}$.

\begin{figure*}
	\includegraphics[width=2\columnwidth]{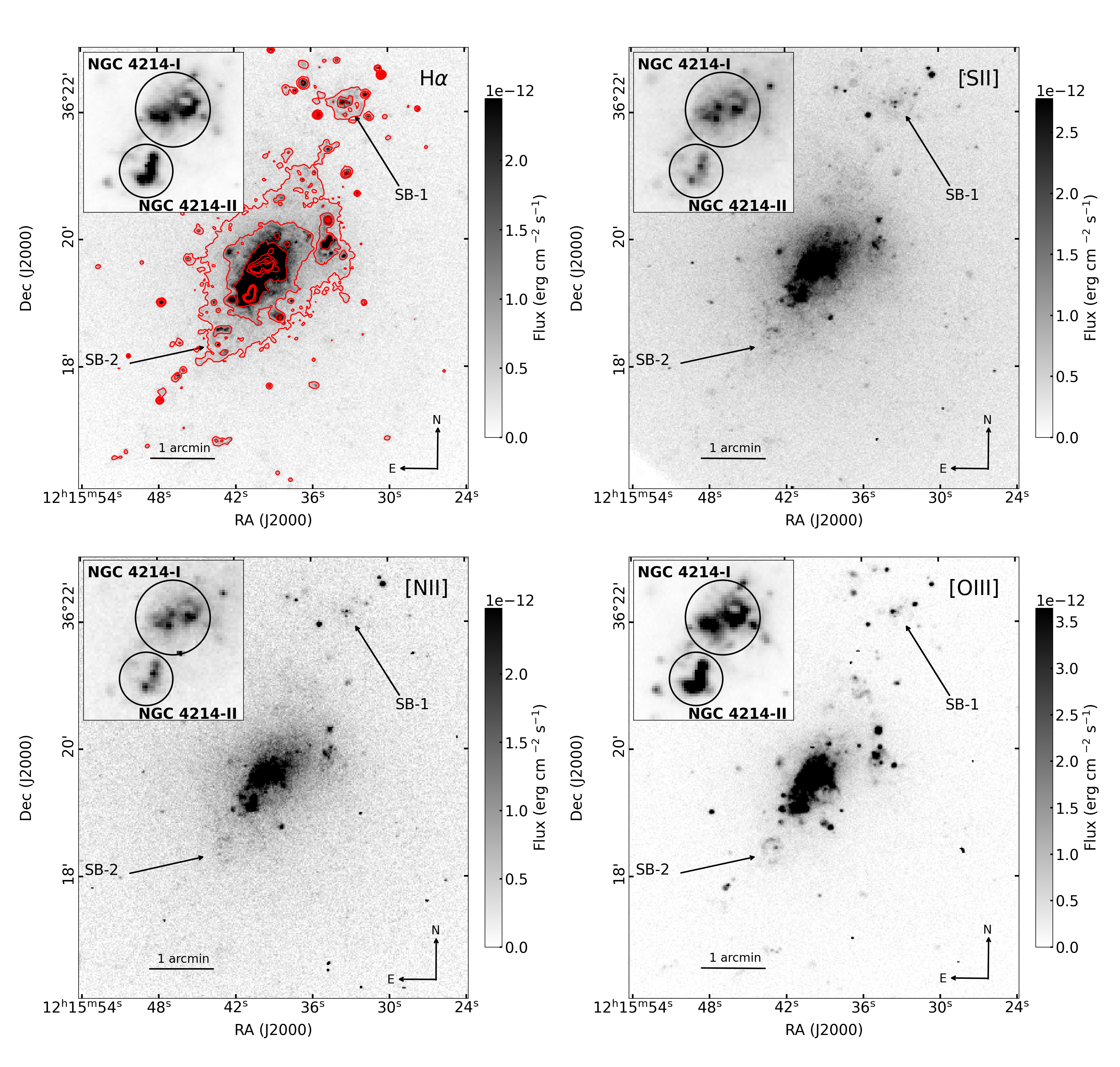}
    \caption{Direct images of NGC\,4214 in H$\alpha$, [\ion{S}{ii}]$\lambda$ 6717 \AA, [\ion{N}{ii}] and [\ion{O}{iii}]$\lambda$ 5007 \AA\ emission lines. The first contour level in the H$\alpha$ image is at 0.41$\times$ 10$^{-12}$ergs cm$^{-2}$ s$^{-1}$(3$\sigma$ over the background). Next contours are at
0.81, 2.32, 8.12, 13.90, and 22.0  $\times$ 10$^{-12}$ ergs cm$^{-2}$ s$^{-1}$. We indicated the position of the superbubbles SB-1 and SB-2. For each direct mage the two large H II complexes NGC\,4214-I and NGC\,4214-II are shown in an inset panel.}
    \label{Fig1}
\end{figure*}

\section{General morphology}
\label{morpho}
The morphology of the ionised gas emission in the central parts of NGC\,4214 has been described by \cite{MacKenty2000, Dopita2010} and \cite{Choi2020}. Here we review the main features given by \citet{MacKenty2000} who described in detail the morphology of the central part of the galaxy and we also describe the global morphology of the ionised gas.

The H$\alpha$ ionised gas distribution of the central part of the galaxy (see inset panel of Fig. \ref{Fig1}) shows the two large H II complexes known as NGC 4214-I and NGC 4214-II \citep{MaizApellaniz1998}. Also, there are a number of isolated fainter knots scattered in the central part of the galaxy and throughout the galaxy field. In Fig. \ref{Fig1} is possible to see some nebulae, filaments and SBs distributed along the galaxy.  In addition, the diffuse interstellar gas (DIG) is surrounding the two main complexes and the isolated knots.

\begin{figure}
	\includegraphics[width=\columnwidth]{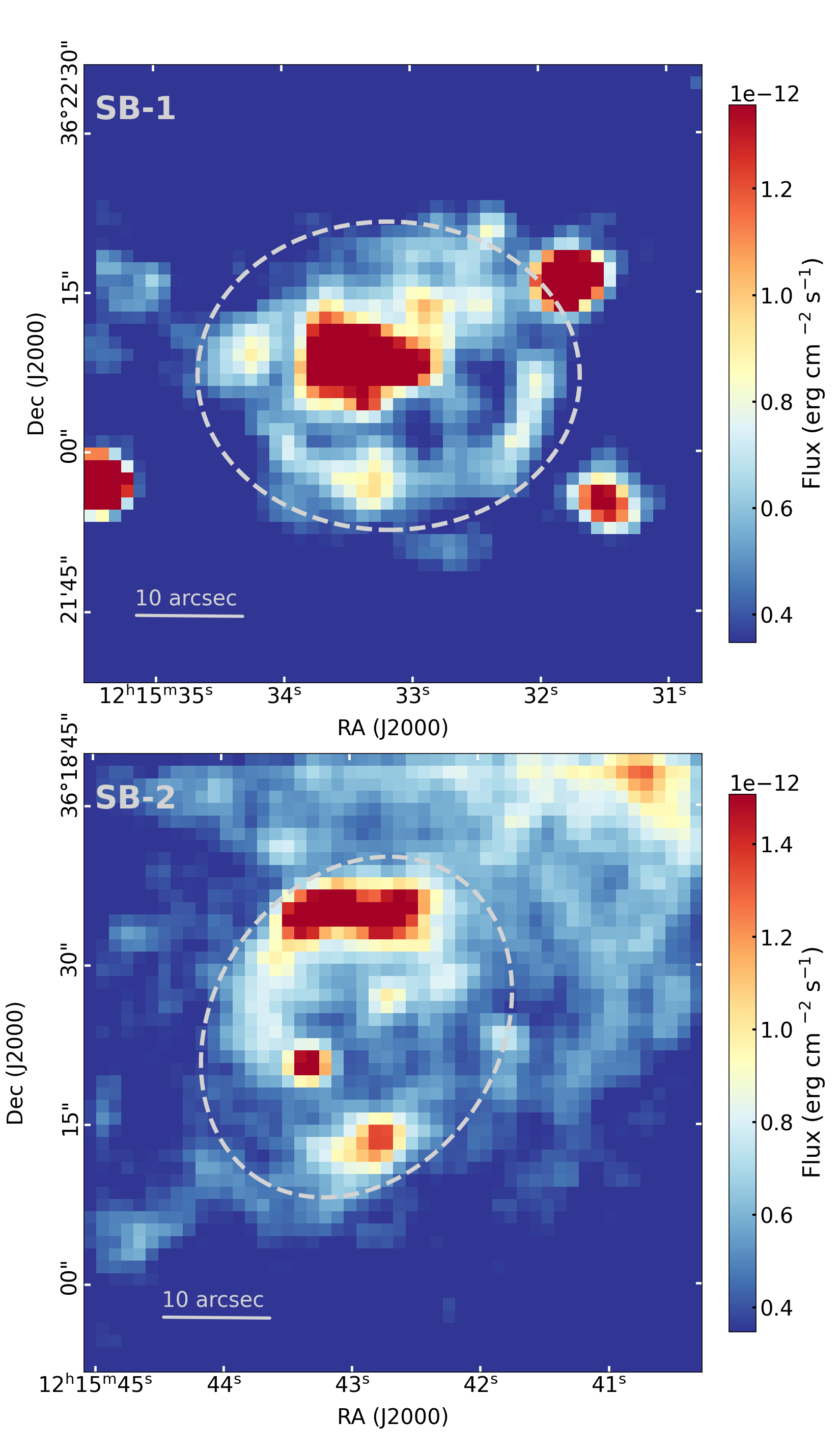}
    \caption{H$\alpha$ images of the SB-1 (top) and SB-2 (bottom) obtained from PUMA. In both images gray dashed ellipses denote the respective SB.}
    \label{Fig2}
\end{figure}

From the direct images we morphologically identified two SBs (named SB-1 and SB-2) located in the galaxy disk and pointed out with arrows in all panels of Fig. \ref{Fig1} and bounded by the contours over plot in the H$\alpha$ image. SB-1 (RA (J2000) =12:15:33.1,  Dec (J2000) =+36:22:06.6) is located at SE edge of the galaxy  and SB-2 (RA (J2000) =12:15:42.9,  Dec (J2000) =+36:18:23.0) is located at 44.5\arcsec (582 pc) of the NGC\,4214-II \ion{H}{ii} complex. The size of the SBs was measured by using the PUMA H${\alpha}$ direct image (see Fig. \ref{Fig2}), the distance to the galaxy (2.7 Mpc) and the image scale (1.16\arcsec).  Figure \ref{Fig2} shows the H${\alpha}$ SBs pointing out the gray dashed ellipses; the semi major and minor axis of the SBs are 36\arcsec$\times$29\arcsec (546.6 pc$\times$440.3 pc) for SB-1 and 34\arcsec $\times$27\arcsec (516.2 pc $\times$ 409.9 pc) for SB-2.

The global morphology of the galaxy in the [\ion{O}{iii}] line is roughly similar to H$\alpha$; i.e., has similarities in the central area and small differences in the galaxy outskirts. Also, the knots situated in the outskirts of the central part are detected in both images. The [\ion{S}{ii}] image shows similar morphology than H$\alpha$ and [\ion{O}{iii}] images while the [\ion{N}{ii}] direct image shows the poorer S/N ratio showing only the central part of the galaxy.

\begin{figure*}
	\includegraphics[width=2\columnwidth]{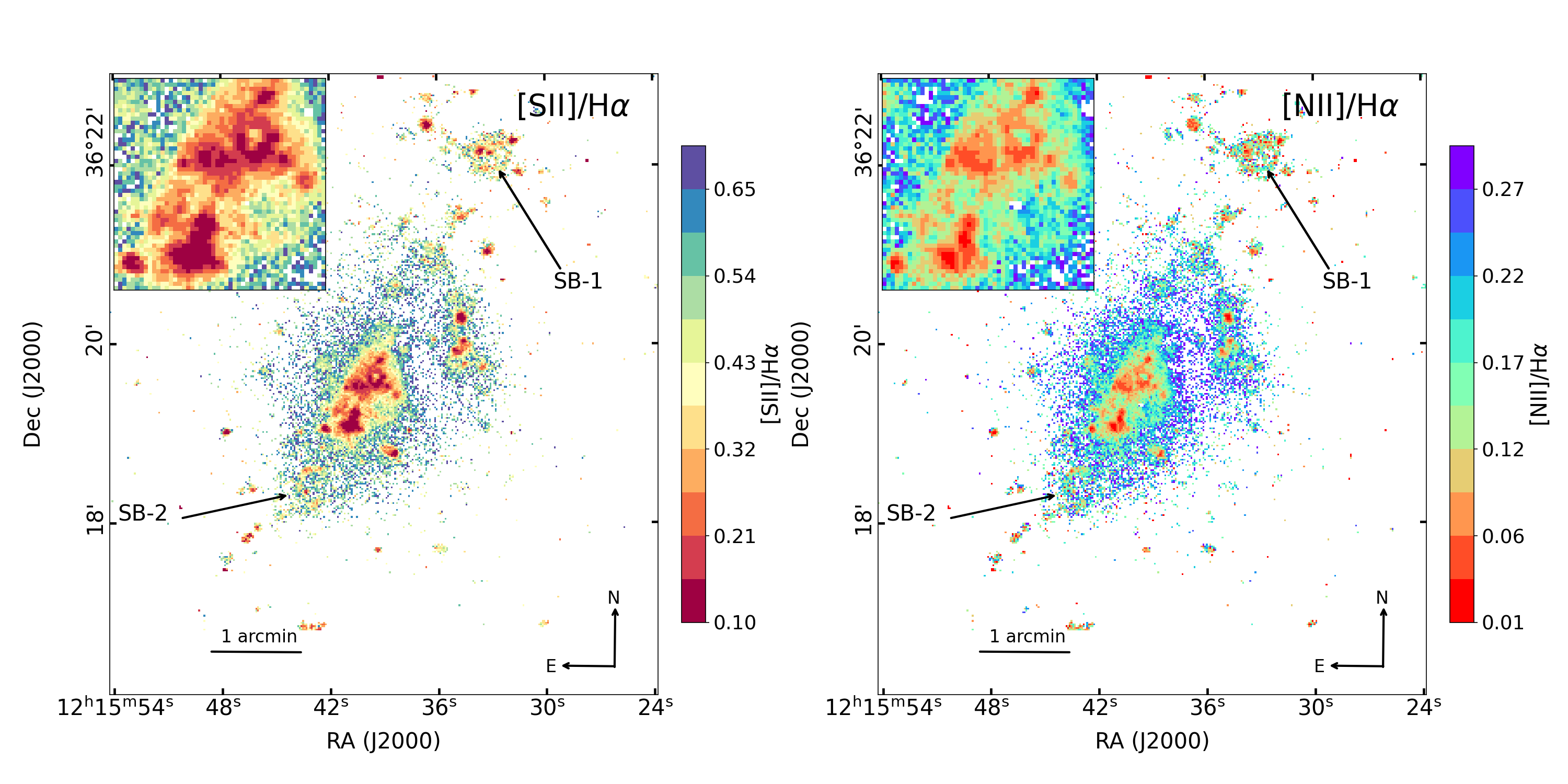}
    \caption{[\ion{S}{ii}]$\lambda$6717/H$\alpha$ and [\ion{N}{ii}]$\lambda$6584/H$\alpha$ line ratios maps of the galaxy NGC\,4214. In both maps an inset panel shows the central part of the galaxy. The black arrows show the position of the two superbubbles (SB-1 and SB-2) found in this work.}
    \label{Fig3}
\end{figure*}

We have also produced maps of the [\ion{S}{ii}]$\lambda$6717/H$\alpha$ and [\ion{N}{ii}]$\lambda$6584/H$\alpha$ line ratios (see Fig \ref{Fig3}); which are close enough in wavelength to be strongly affected by dust attenuation. Those maps allow us to study the ionised structure of the galaxy.
Low values of [\ion{S}{ii}]$\lambda$6717/H$\alpha$ and [\ion{N}{ii}]$\lambda$6584/H$\alpha$ are located in the central part of the galaxy, western region and in SBs, corresponding to the high excitation zone. The 
higher ratio values ([\ion{S}{ii}]/H$\alpha \sim$ 0.65) are located at the outskirts of the galaxy (where the DIG is located) depicting the low excitation regions.

When we compared our excitation ratio maps of the two large H II complexes (see the inset panels of Fig. \ref{Fig3}) with those obtained by \citep[][see their figures 7 and 9]{MaizApellaniz1998}, we found an excellent agreement with values of 0.1<[\ion{S}{ii}]/H$\alpha$<0.25 and  0.07<[\ion{N}{ii}]/H$\alpha$<0.12. 
 
The values of [\ion{S}{ii}]/H$\alpha$ for the SBs are between 0.1 and 0.3 while for [\ion{N}{ii}]/H$\alpha$ are between 0.01 and 0.12. Both line ratio values are located mainly in the shell of the SBs. 
Our [\ion{S}{ii}]/H$\alpha$ are similar to the values of SBs in the Large Magellanic Cloud \citep[ranges from 0.0 to 0.6,][]{Georgelin1983} and in NGC\,1569  \citep[ranges from 0.06 to 0.44,][]{SanchezCruces2015}.

The kinematic analysis of the SBs will be discussed later in Section \ref{SBs}.

\section{Kinematics of the Ionised gas}

\subsection{Velocity field}
\begin{figure*}
	\includegraphics[width=2\columnwidth]{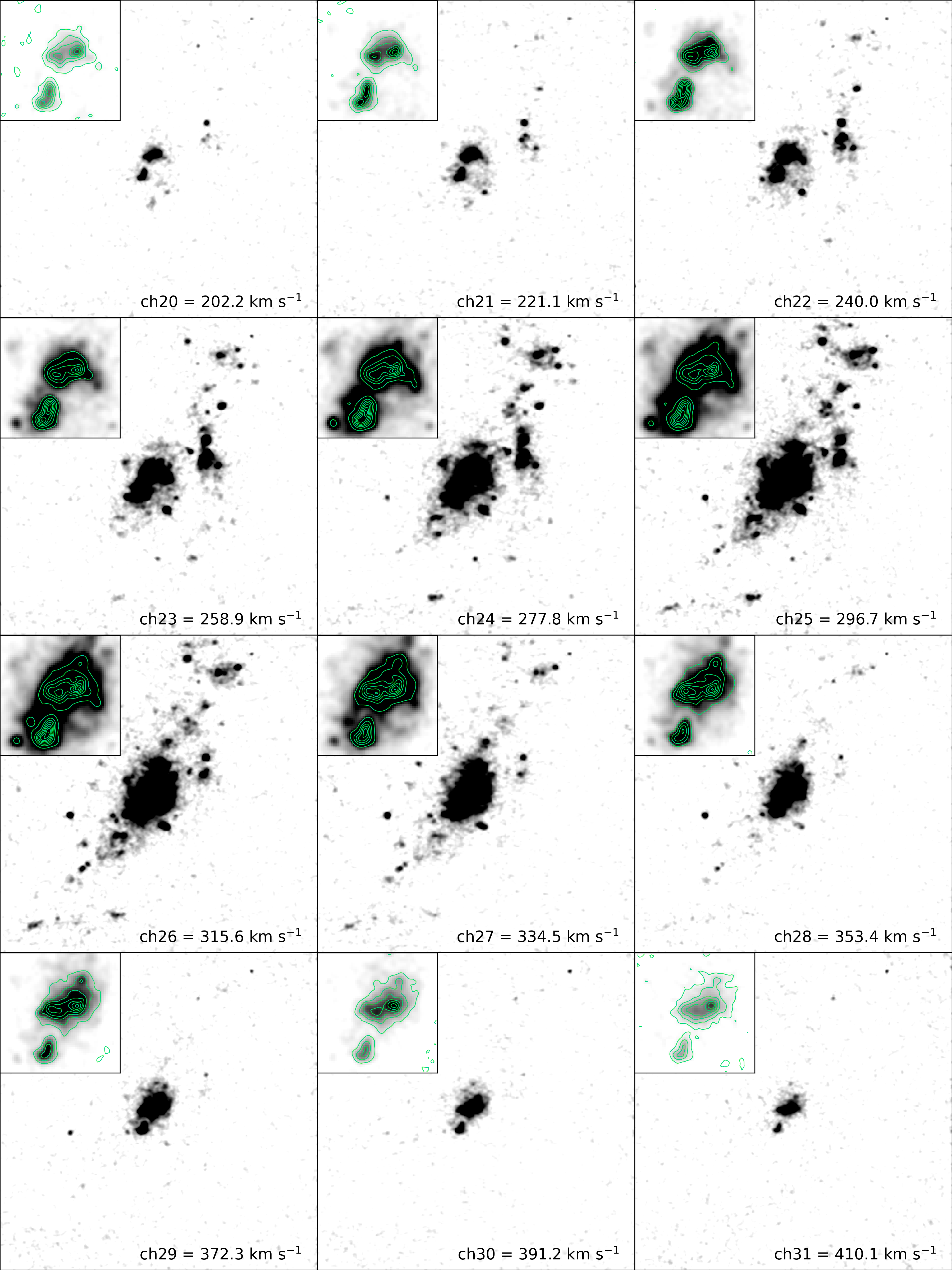}
    \caption{Heliocentric velocity channels of galaxy NGC\,4214 with the continuum subtracted. The velocity of each channel is shown in bottom right. The inset panels show a close-up of the central part of the galaxy in their respective velocity channels.  Contour levels were automatically chosen between minimum and maximum intensity values in order to show the morphology of the region.}
    \label{Fig4}
\end{figure*}

Figure \ref{Fig4} shows some velocity channels (obtained from the velocity data cube) of the H$\alpha$ emission detected in NGC\,4214; velocities range from 202 to 410  km\,s$^{-1}$.  Also in this figure we show in each channel map a close-up of the central part of the galaxy with contours  superimposed. Contour levels were automatically chosen between minimum and maximum intensity values in order to show the morphology of the region in the respective velocity channel. In general, the contours show the morphology on the  H II complexes NGC\,4214-I and NGC\,4214-II. The receding velocities of  NGC\,4214-II are at the northwest and the approaching are at southeast. However for NGC\,4214-I it is not easy to see a velocity pattern.

\begin{figure*}
\includegraphics[width=2\columnwidth]{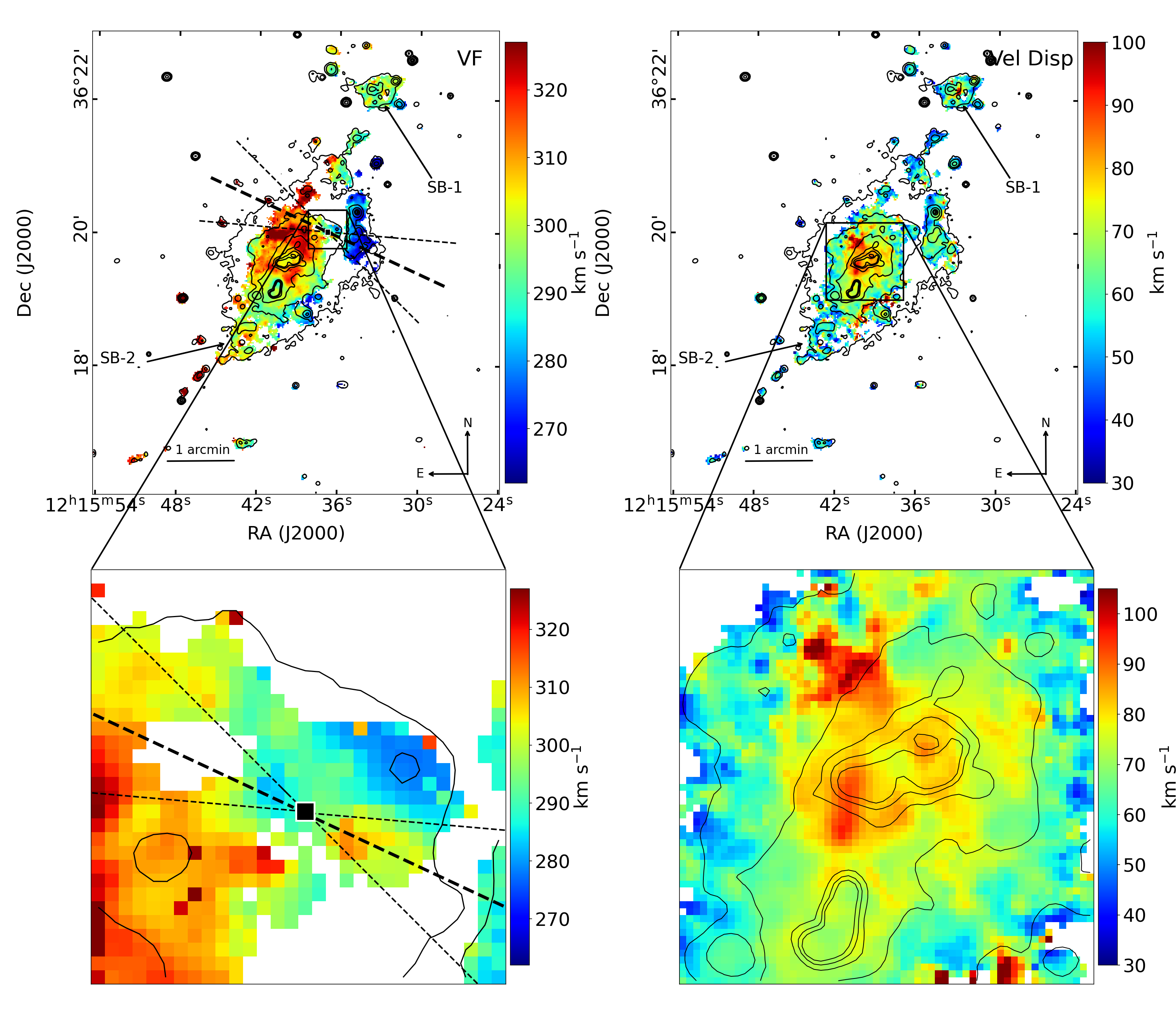}
\caption{\textit{Left panels}: H$\alpha$ velocity map of NGC\,4214 (top) and close-up of the central parts of the galaxy (bottom) with the kinematical centre ('$\blacksquare$') overplotted. The thick dashed line represents the kinematic major axis and the thin dashed lines represent the angular sector (20$\degr$) used to derive the rotation curve. \textit{Right panels}: Velocity dispersion field (top) and a close-up of the \ion{H}{II} complex. The contours on both fields indicate the flux of the H$\alpha$ direct image (Fig. \ref{Fig1}).}
\label{Fig5}
\end{figure*}

Figure \ref{Fig5} shows the velocity field (VF) of NGC\,4214. From this figure we can see that the galaxy does not follow  circular rotation, no rotation axis can be found and  the gas motions are disordered. This is consistent with previous studies of \citet{Garrido2004} and \citet{Epinat2008}. The VF displays velocity values ranging from 250 to 350 km s$^{-1}$ (implying a difference of about V$\sim$100 km s$^{-1}$); this difference of values are located mainly at the northwest (NW) side of the galaxy. While, the southeast (SE) side of the galaxy presents almost constant velocities (at about 300 - 330 km\,s$^{-1}$). The velocities of the two large H II complexes are about 300 km\,s$^{-1}$ for NGC\,4214-I  and 290-300 km\,s$^{-1}$ for NGC\,4214-II. The VF highlights that NGC\,4214-I is surrounded by the highest velocity values at about 320-350 km\,s$^{-1}$. At the NW of the galaxy, there is a high blueshifted region velocities about V = 270-290 km\,s$^{-1}$.

In general, the velocity dispersion map (right hand panel of Fig. \ref{Fig5}) shows values from 30 - 80  km\,s$^{-1}$ except for the central part, where the high dispersion velocity values (between 90 and 110 km\,s$^{-1}$) are distributed mainly along NGC\,4214-I and are in excellent agreement with those determined in similar studies of starburst dwarfs \citep[e.g.][]{Marasco2023}.  In the center of the high dispersion region, there is a small region with $\sim$75 km\,s$^{-1}$ confirming the existence of an expanding shell related to NGC\,4214-I. 

\begin{figure}
\includegraphics[width=1\columnwidth]{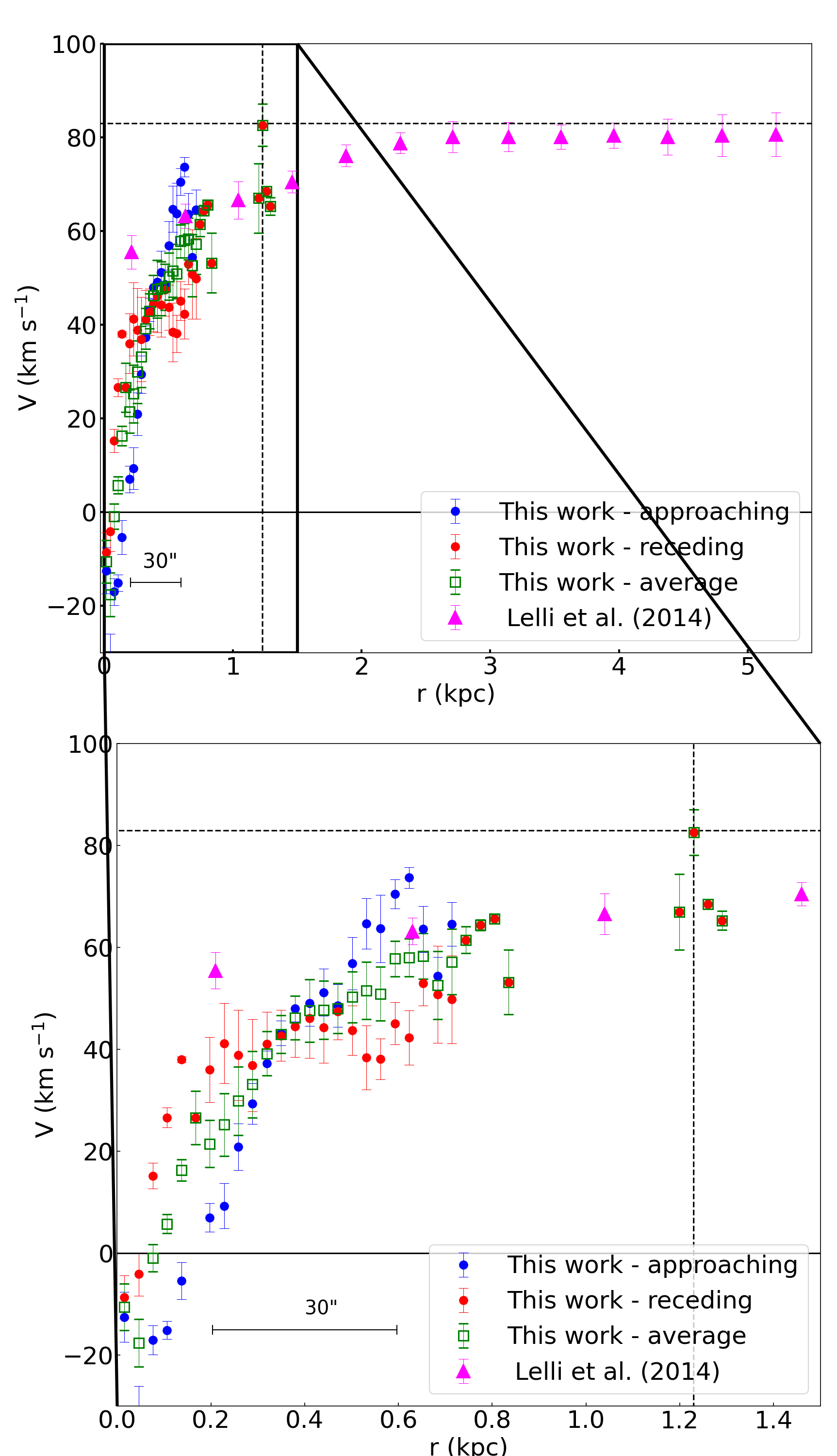}
\caption{Rotation curve of the galaxy derived from the H$\alpha$ velocity field. In the bottom panel we show a close up of the RV at 1.5 kpc. In both panels, the receding side corresponds (red circles) to the western side of the galaxy and the approaching side (blue circles) to the eastern side. The average of the receding and approaching sides is indicated with green open squares. Pink triangles correspond to the \ion{H}{I} RC derived by \citep{Lelli2014} with a resolution of (30 arcsec).}
\label{Fig6}
\end{figure}

\subsection{Rotation curve}

Previous works of \cite{Garrido2004} and \cite{Epinat2008} using H$\alpha$ Fabry-Perot observations showed that this galaxy does not present any evidence for rotation in its velocity map; hence the H$\alpha$ rotation curve could not be computed.  However, \cite{Swaters2009} computed the rotation curve of NGC\,4214 as part of a sample of 62 galaxies from the \ion{H}{I} Survey of Spiral and Irregular Galaxies (WHISP) project. Later, \cite{Lelli2014} using radio observations of the VLA, computed the rotation curve and made the asymmetric-drift correction and confirmed that this galaxy shows non-circular motions in its disk likely associated with the \ion{H}{I} spiral arms and/or the inner stellar bar. This bar is located at the place of the two large H II regions and coincides with the galaxy’s minor axis of the \ion{H}{I} map \citep[see][]{Walter2008, Lelli2014}. 
\cite{Lelli2014} gave two set of kinematical parameters  (position angle, PA and inclination, i) for NGC\,4214 due to its \ion{H}{I} disk is close to face-on in the inner parts and present strongly warped in the outer parts. They found, a PA$_\text{kin-\ion{H}{I}}\simeq$65$\degr$ and i$_\text{kin-\ion{H}{I}}\simeq$ 30$\degr$ for R $\lesssim$ 3$\arcmin$, and PA$_\text{kin-\ion{H}{I}}\simeq$ 84$\degr$ and i$_\text{kin-\ion{H}{I}}$ gradually decreasing at larger radii and, that the optical and \ion{H}{I} spiral arms wind in opposite directions, clockwise and counter-clockwise, respectively. 

In this section, we attempted to derive a RC from the H$\alpha$ VF of the galaxy following the methodology in \cite{Amram1996}, \cite{FuentesCarrera2004} and \cite{SanchezCruces2022} taking into consideration the kinematical parameters given by \cite{Lelli2014} for R $\lesssim$ 3$\arcmin$.

The RC was obtained with the ADHOCw software using the standard tilted-ring method. In this case, we considered a width of three-pixels (45 pc). As we discussed in \cite{SanchezCruces2022}, to this method be valid the gas has to move in purely circular orbits, which as mentioned before, is not  the case for the innermost parts of NGC\,4214; therefore, we derived its RC considering radii larger than 200 pc up to 1.51 kpc (100 pix) within an angular sector of 20$\degr$ along the galaxy major axis (in \ion{H}{I}). 

To compute the kinematics parameters with ADHOCw, start with first guess values for \textit{i}, PA, centre $(x_0, y_0)$ and V$_{sys}$, derived from the literature; then those parameters were iteratively modified to obtain a symmetry between the approaching and receding sides superposed and making sure that the RC had low dispersion values \citep[e.g.][]{Amram1992,Amram1996,FuentesCarrera2004,SanchezCruces2022}. In this case we used as initial parameters those of \cite{Lelli2014}; kinematical centre (RA$_\text{kin-\ion{H}{I}}$(J2000)=12:15:36.9, Dec$_\text{kin-\ion{H}{I}}$ (J2000)=+36:19:59.0), \textit{i}=30\degr and a systemic velocity V$_\text{sys}$ =290 km s$^{-1}$.

The derived RC is shown in Fig. \ref{Fig6} and the kinematical parameters are PA=63\degr, V=293 km s$^{-1}$ and \textit{i}= 30\degr; in this case we used the same kinematical centre of \cite{Lelli2014}. From Fig. \ref{Fig6} we can see that in the innermost region of the RC (radii smaller than 50 pc), the rotation velocity values of the approaching and receding sides are separated by $\sim$5 km s$^{-1}$, with the rotation velocity of the receding side reaching $\sim$-8 km s$^{-1}$ while the rotation velocity on the approaching side reaches $\sim$-13 km s$^{-1}$. Beyond this radius (R~=~50~pc), the difference in velocities between the receding and approaching side increases (differences about 30 km s$^{-1}$) until reaching a radius of 0.3 kpc. Subsequently, between 0.3 and 0.5 kpc, the RC is symmetric with the smaller dispersion; after this radius the RC is no longer symmetric and with oscillations, rotation velocity differences between 20 and 30 km s$^{-1}$ and displaying an ascending behaviour with a maximum rotation velocity of about 82 km s$^{-1}$ at R = 1.23 kpc.

In Fig.\ref{Fig6} we also show the comparison between H$\alpha$ RC derived from our PUMA observations with the \ion{H}{I} RC obtained by \cite{Lelli2014}. We can see that the H$\alpha$ RC extended up to 1.4 kpc while the \ion{H}{I} RC extended beyond 5 kpc. There are only a few points of the \ion{H}{I} RC  within the H$\alpha$ RC due to the spatial resolution difference between \ion{H}{I} (30 arcsec) and H$\alpha$ data (1.16 arcsec). Three \ion{H}{I} points at 0.6, 1.04 and 1.46 kpc are compatible within twice the quoted rms uncertainty H$\alpha$ data.
The differences between RCs is due to NGC\,4214 is very close to face-on and has a strong warp \citep{Lelli2014}. Also, 
the asymmetric behaviour in the innermost parts of the galaxy (R<0.2 kpc) of the H$\alpha$ RC is an indication that the velocity dispersion of the ionised gas in NGC\,4214 is mainly determined by the energy injected into the ISM by ongoing SF \citep[see][]{Moiseev2015}.

\subsection{Evidence for Expanding Superbubbles}
\label{SBs}

\begin{figure*}
	\includegraphics[width=2\columnwidth]{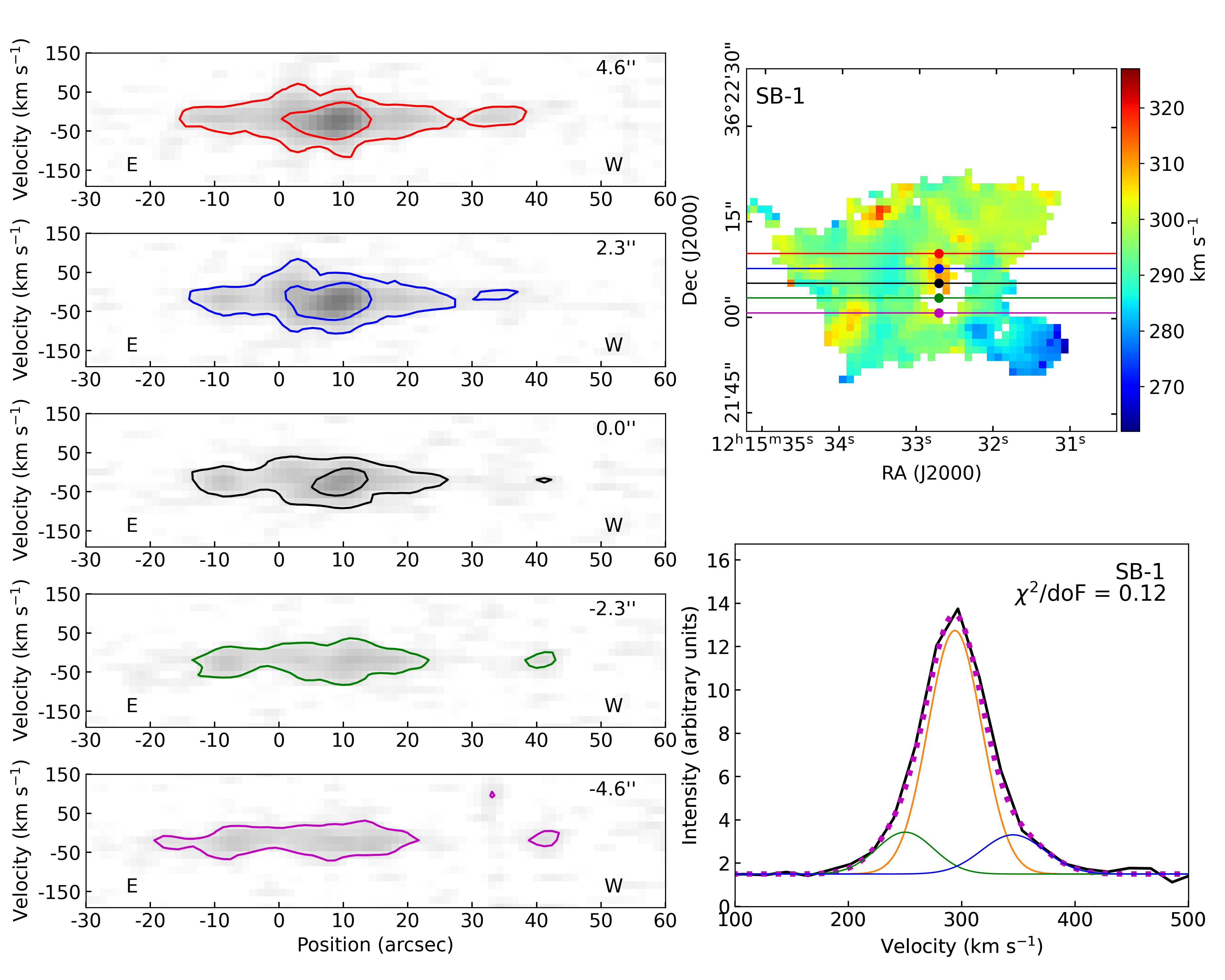}
    \caption{Left: Position Velocity Diagrams (PVDs) of SB-1 of five pseudo-slit positions on the velocity field of SB-1 in Top-right panel. The black slit corresponds to that one in their geometric centre and the other four pseudo-slit are distributed parallel and equidistant with a 2.6 $\arcsec$ of separation between them. Bottom-right is the SB-1 velocity profile fitted with three velocity components shown in orange for the brighter one, green for the blueshifted component and blue for the redshifted component.}
    \label{Fig7}
\end{figure*}

\begin{figure*}
	\includegraphics[width=2\columnwidth]{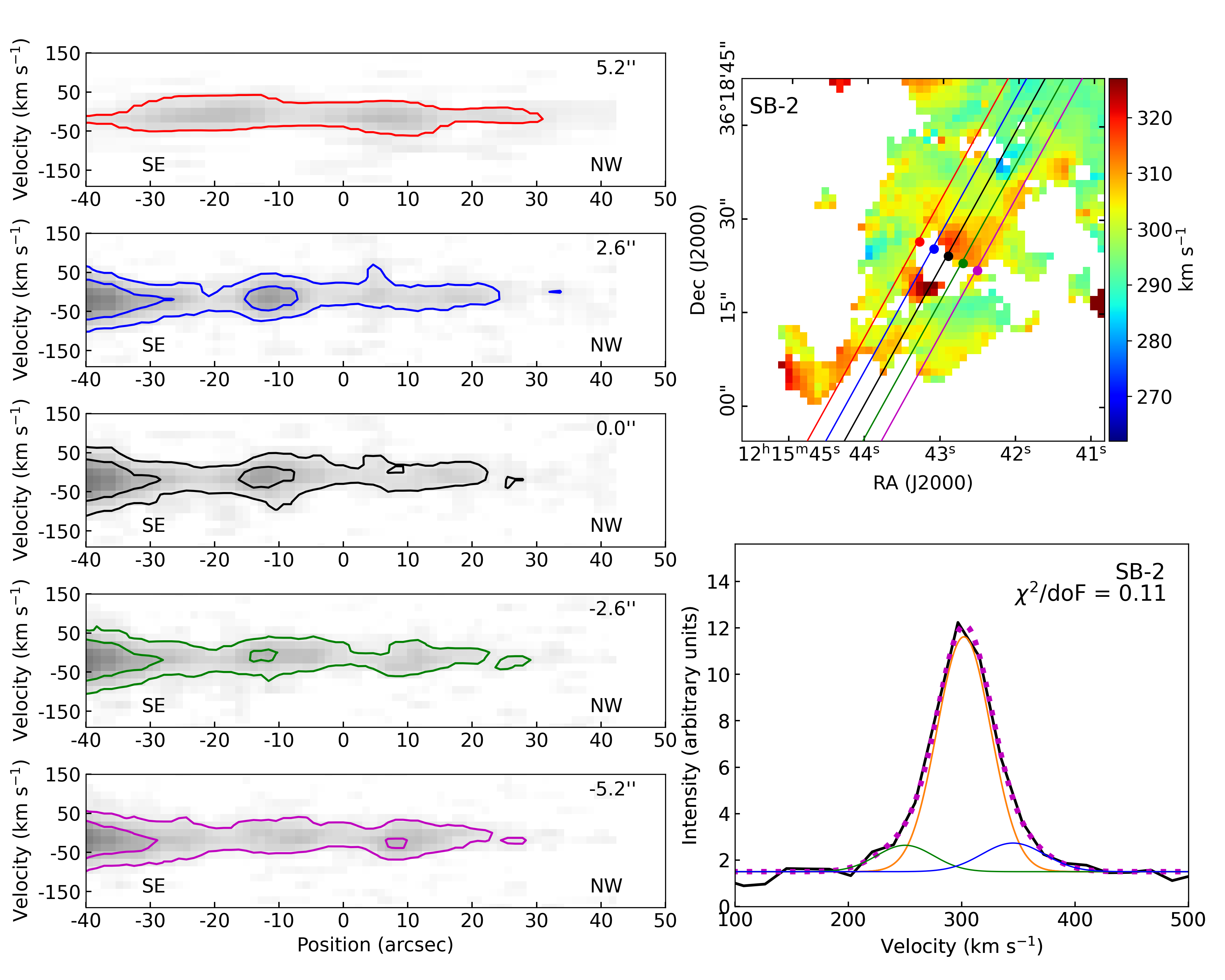}
    \caption{Similar to Fig. \ref{Fig7} for SB-2. }
    \label{Fig8}
\end{figure*}

The VF of the SB-1 and SB-2 are shown in top right-hand panels of Fig. \ref{Fig7} and \ref{Fig8}, respectively. The velocities inside the SB-1 are about 287 km\,s$^{-1}$ and on the edge of the shell there are three knots with velocities $>$300 km\,s$^{-1}$.
For SB-2, the highest velocities are coming from the central part  with velocities $>$310 km\,s$^{-1}$ while  velocities of the shell are about 295 km\,s$^{-1}$. 

We extract the velocity profiles of the SBs over square boxes encompassing the whole extent of each superbubble. As we can see in the right-bottom panels of Fig. \ref{Fig7} and \ref{Fig8}, the velocity profile of both SBs presents wings or humps, indicating the presence of composite profiles. The decomposition of the profiles were constrained by fitting the minimum number of Gaussian functions by using a combination of visual inspection and the $\chi^2$/doF value of the fit; i.e., when we fitted a profile with a single Gaussian and visually this fit did not match the observed velocity profile; we then fitted it with two or three components (depending on the profile) plus checking the $\chi^2$/doF obtained in each fit \citep[see][]{SanchezCruces2022} which indicates the accuracy of the fit.

\begin{figure}
	\includegraphics[width=\columnwidth]{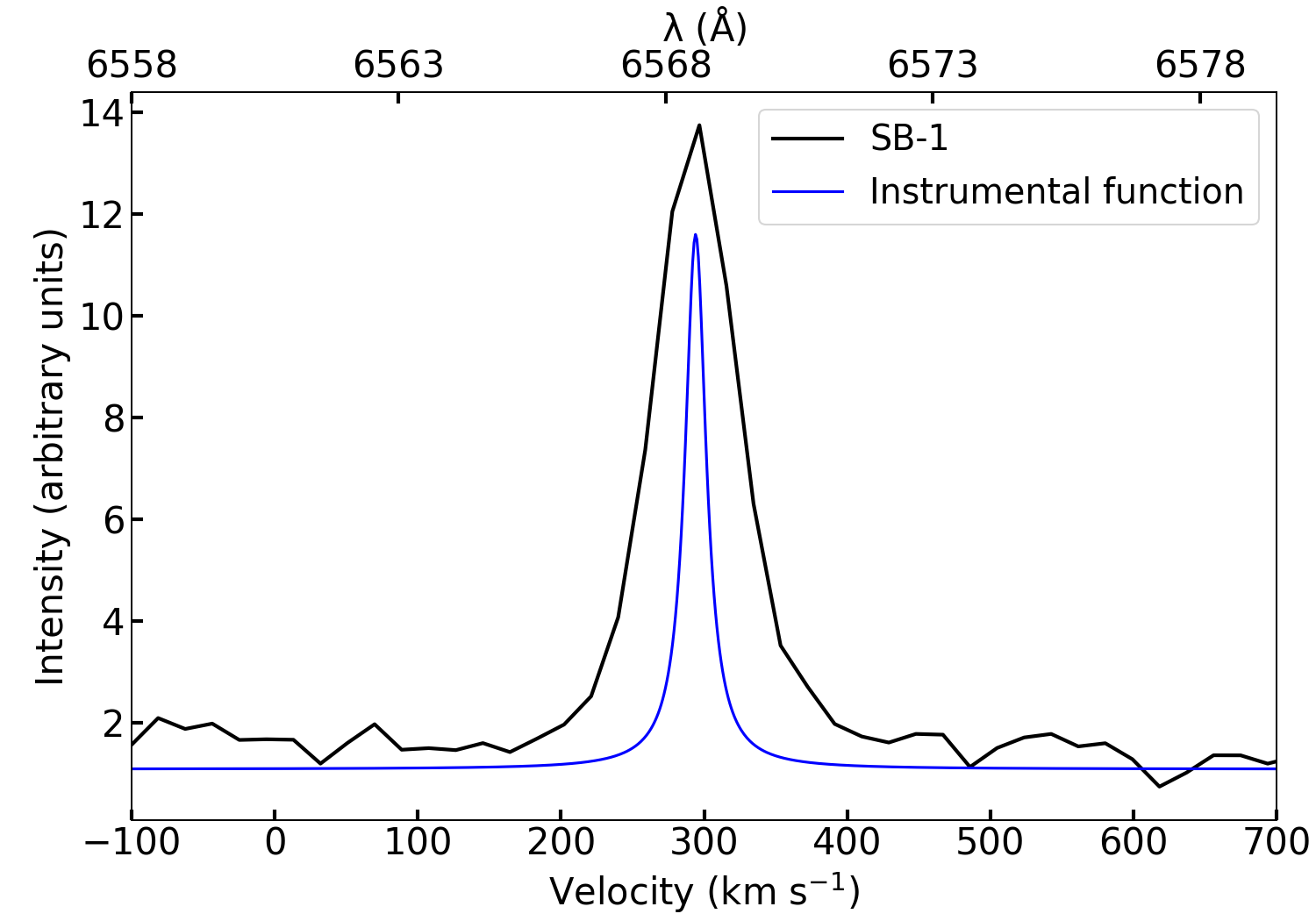}
    \caption{Comparison between H$\alpha$ velocity profile of SB-1 (black line) and instrumental function (Airy function, blue line).}
    \label{Fig9}
\end{figure}

The comparison between velocity profiles of a HII region fitted with a single Gaussian component and complex velocity profiles of shock nebula fitted with two or three Gaussian components was presented in \cite{SanchezCruces2022}. Additionally, here we show in Fig. \ref{Fig9} the instrumental contribution overplotted in the velocity profile of SB-1. The instrumental profile is represented by the Airy Function (Lorentzian function). The velocity profile observed in each pixel  is the convolution of the Airy function and the 'intrinsic' emission line. Although this function minimally affects the central value and FWHM of the velocity profile, this is calculated with the baricenter of the profile, and as we can see in Fig. \ref{Fig9} the shape of the instrumental function does not introduce high velocity wings in the velocity profile.

The velocity profile of both SBs were fitted with three Gaussian functions; we interpreted the main component (in orange) as the 'systemic' speed of the SB associated with its rotational motion around the galactic centre. The blueshifted and redshifted components are presented in magenta and cyan respectively, and are related to the expansion velocity of the SBs. For SB-1 the main component is  294 km\,s$^{-1}$ and for SB-2 is 302  km\,s$^{-1}$, while the second components are 250 km\,s$^{-1}$ (blueshifted component) and 345 km\,s$^{-1}$ (redshifted component) for both SBs. Besides, when we convolve the Gaussian functions with the Airy Function, we realized that there are no significative changes in the position of the velocity components of their profile indicating that the instrumental broadening has a negligible impact on the velocity profiles.

In order to support the results obtained in the velocity profiles fitted, we show in Figs \ref{Fig7} and \ref{Fig8} the Position Velocity Diagram (PVDs) of each SB centered in their respective geometric centre; these PVDs were extracted from the FP data cube after having subtracted the stellar continuum and along each SB. For SB-1 the pseudo-slit positions used to extract the PVD are from East to West (see Fig. \ref{Fig7}) while for SB-2 the pseudo-slit positions are from Northwest to Southeast (see Fig. \ref{Fig8}). For both PVDs, the black slit corresponds to that one in their geometric centre and the other four pseudo-slit are distributed parallel and equidistant with a 2.6 $\arcsec$ of separation between them. Also, in both PVDs (left panels of Figures \ref{Fig7} and \ref{Fig8}) it is possible to see half of the so-called Doppler ellipse of the SBs \citep[see also][]{Lozinskaya2003} corresponding to the approaching side of the shell and the velocity variation with the SB center associated with the approaching and receding sides of the SBs. On the other hand, S/N ratio of the secondary Gaussian components of the SBs fit (see Fig. \ref{Fig7} and \ref{Fig8}) are marginal, about 2, indicating that these components are faint but are observed in both, the velocity profiles and in the PVDs.

For SB-1, the evidence of the Doppler ellipse is more clear in the -2.3 and -4.6 arcsec PVDs position where half of the called  Doppler ellipse corresponds to the approaching side of the shell. The PVD at +2.3 
exhibits a large velocity variation (from -100 to +100 km\,s$^{-1}$) at position between 3 and 6 arcsec; the PVD at +4.6 present two large variation in velocity of -100 to 90 km\,s$^{-1}$ (at 5 arcsec) and -105 to 80 km\,s$^{-1}$ (at 10 arcsec). For SB-2, the PDV at the centre position (black contours) shows half feature of the Doppler ellipse corresponding to the approaching side of the shell even a second half Doppler ellipse is seen between -30 to -10 arcsec, 
which is not a part of SB-2. PVD at -5.2 shows no clear (week) signs of expansion. The PVD at -2.6 present an arc structure between -10 and 10 arcsec related with the half of the velocity ellipse. In the +2.6 and +5.5 positions the arc structure is barely seen.

\subsubsection{Expansion Velocity}

The expansion velocities of the SBs were obtained considering the expansive motion of a shell as V$_\text{exp}$ = (V$_\text{max}$-V$_\text{min}$ )/2 \citep[see][]{AmbrocioCruz2004, SanchezCruces2015}, where Vmax and Vmin are the redshifted and blueshifted velocity components, respectively. In this case both SBs have the same velocity expansion of V$_\text{exp}$ = 47.5$\pm$2 km\,s$^{-1}$, typical velocity expansion for SB \citep[see][]{Roy1991,Weiss1999,ValdezGutierrez2001,Lozinskaya2008,SanchezCruces2015,Egorov2021}.

\subsubsection{Physical Parameters}
The physical parameters of the SBs (rms electron density (n$_\text{e}$), the mechanical energy (E$_\text{mec}$) of the stellar winds, and the kinematic age (t)) were computed following \citet{ValdezGutierrez2001} and \citet{SanchezCruces2015} works. The n$_\text{e}$ was obtained with the relation: 

\begin{equation}
\label{eq1}
n^2_e (rms)(cm^{-6} ) = 2.74 \times 10^{18} F(H\alpha)\theta^{-2} R^{-1}
\end{equation}

where F(H$\alpha$) is the H$\alpha$ flux distributed in a spherical shell of radius R (in parsecs) and $\theta$ is the semiaxis size of the SB  in arcsecond units.

The mechanical energy of the superbubbles can be derived from their expansion velocity (E$_\text{mec}$ = 1/2 MV$_\text{exp}$) and assuming that the mass of the SB shell is concentrated in a spherical shell of
radius R (in parsecs) and thickness $\Delta$R= R/12 and considering a $\mu$ = 0.65 for an ionised hydrogen gas \citep[see][]{ValdezGutierrez2001,SanchezCruces2015}. Thus, the mechanical energy in the ionised superbubble shell is E$_\text{mec}$ (erg)= 1/2 MV$_\text{exp}$ = $1.7 \times 10^{41} \text{n}_\text{e}(\text{cm}^{-3}) \text{V}_\text{exp}^2 \text{R}^3$.

For the kinematic age (t), we considered the energy-conserving model \citep{McCray1977, Weaver1977}:

\begin{equation}
\label{eq2}
t(yr)= 4.8 \times 10^5 \text{R/V}_\text{exp}
\end{equation}

where R is the radius and V$_\text{exp}$ is the expansion velocity of the SB.

We found that the SB-1 has a rms electron density n$_\text{e, SB-1}$~=~2.33$\pm$1.30 $\text{cm}^{-3}$, a  mechanical energy E$_\text{mec,SB-1}$~=~9.55$\pm$5.40 $\times$10$^{51}$ erg and a kinematic age t$_\text{SB-1}$~=~2.22$\pm$0.09 Myr. For SB-2 we found a  n$_\text{e,SB-2}$~=~2.63$\pm$0.88 $\text{cm}^{-3}$, an E$_\text{mec,SB-2}$~=~8.68$\pm$2.99 $\times$10$^{51}$ erg and a kinematic age t$_\text{SB-2}$~=~2.07$\pm$0.09 Myr.

As we see, both SBs are relatively young with ages of about $\sim$2 Myr. Those values are consistent with the ages estimated in  IC\,1613 \citep[0.7-2.0,][]{ValdezGutierrez2001},  NGC\,1569 \cite[0.5 -3.1,][]{SanchezCruces2015}  and DDO\,53 \cite[1.1-2.9][]{Egorov2021}. The energy value we obtained for SBs is similar to the SBs obtained in IC\,1613, NGC\,1569 and DDO\,53 with values between $\sim$0.1 and $\sim$10$^{51}$ erg \citep[see][]{Meaburn1980,Bruhweiler1980,ValdezGutierrez2001,Bullejos-Rosado2002, Thurow-Wilcots2005,Lozinskaya2008, SanchezCruces2015, CampsFarina2017,Ramachandran2018,Egorov2021}. 

One of the most feasible scenarios to better explain the mechanism for the formation of a SB is the constant energy injection from multiple stars plus supernovae \citep{McCray1987,MacLow1988,TenorioTagle1988, MacLow1989}. Due to the fact that the superbubbles physically are very similar to wind-blown interstellar bubbles around individual massive stars, we can consider  the standard model of wind-blown bubbles \citep{Castor1975, Weaver1977}, to describe the SBs in NGC\,4214. The standard model  considers that an OB-type star at rest releases a constant mechanical wind luminosity L$_\text{W}\approx$10$^{36}$ erg s$^{-1}$, with a mass loss rate $\dot{M}_\text{W} \approx$10$^{-6}$ M$_{\odot}$ yr$^{-1}$ and wind terminal velocity V$_W \approx$2000 km s$^{-1}$. 

To compute the mechanical wind luminosity, L${_W}$ of the SBs, we can consider the standard model of wind-blown bubbles \citep{Weaver1977}  given by:

\begin{equation}
\label{eq3}
L_\text{W} (36) = 3.2 \times 10^{-7} n_{0} V_\text{exp}^{3} R^{2}
\end{equation}

where, $\text{n}_{0}$ is the ambient pre-shock density given by $\text{n}_{0} = \text{n}_\text{e}/4$ in units of cm$^{-3}$; considering that the swept-up mass in the shell comes from a homogeneous sphere of radius R in pc, the expansion velocity V$_\text{exp}^{3}$ in km s$^{-1}$, and $\text{L}_\text{W} (36)$ is in units of $10^{36}$ erg s$^{-1}$.  We found that the L$_\text{W}$ for SB-1 is L$_\text{W, SB-1}$ = 9.70$\pm$5.4$\times$10$^{38}$ erg s$^{-1}$ and for SB-2 L$_\text{W,SB-2}$ = 9.47$\pm$3.1$\times$10$^{38}$ erg s$^{-1}$. Table \ref{Table3} shows the derived parameters of the SBs.

The L${_W}$ we obtained for both SBs are in agreement with the L${_W}$ value obtained in the N206 complex in the Large Magellanic Cloud by \cite{Ramachandran2018}; they found that the total mechanical luminosity of N206 is about L${_W}$=1.7 $\times$10$^{38}$ erg s$^{-1}$ generated by the stellar winds of two Wolf-Rayet (WR) stars plus 164 stars in the interior OB association in which the two WR winds alone release about as much wind luminosity (L$_{W,W-R}$ =7.54 $\times$10$^{37}$ erg s$^{-1}$) as the whole in the OB association (L$_{W,OB}$ =8.82 $\times$10$^{37}$ erg s$^{-1}$).

Since N206 complex has a similar size (465 pc) as the SBs in this work and has an expansion velocity of $\sim$20-30 km s$^{-1}$ \citep{Rosado1982,Dunne2001}, we can consider that SB-1 and SB-2 of NGC 4214 could have similar number of OB stars as N206.

Given the dynamical ages of the SBs (2-3 Myr) are shorter than the main-sequence lifetime ($\sim$4-8 Myr) of a 20-60 M$\sun$ stars, the most probable scenario to explain the SBs formation is that only massive stars in OB associations are responsible to form them; i.e., probably non star have had the time to evolve into supernova. This is supported by the fact that if supernovae would have taken place, it would be signs of shock excitation, that is not the case due to the [\ion{S}{ii}]/H$\alpha$ for the SBs are between 0.1 and 0.3. Also, so far there is no information from X-ray emission (with T>10$^6$ K) or non-thermal radio sources (with spectral indices values between -0.2 and -0.7) around the position of the SBs to indicate the presence of shocks.

\subsection{Kinematics and Physical Parameters of SNRs} 
\label{Param_SNRs}

As we mentioned in Section \ref{intro}, NGC\,4214 harbors 35 SNRs detected in radio, optical and X-ray emission; some of them present emission in two or three wavelength bands. From the 35 SNRs only 5 (SNR-2, SNR-3, SNR-5, SNR-23 and SNR-33) present emission in the three wavelengths, 7 present emission in radio and optical (SNR-1, SNR-4, SNR-21, SNR-31, SNR-32, SNR-34 and SNR-35) and 5 (SNR-24, SNR-26, SNR-28, SNR-29 and SNR-30) present emission in optical and X-ray (see Table \ref{Table4}).

\begin{figure}
	\includegraphics[width=\columnwidth]{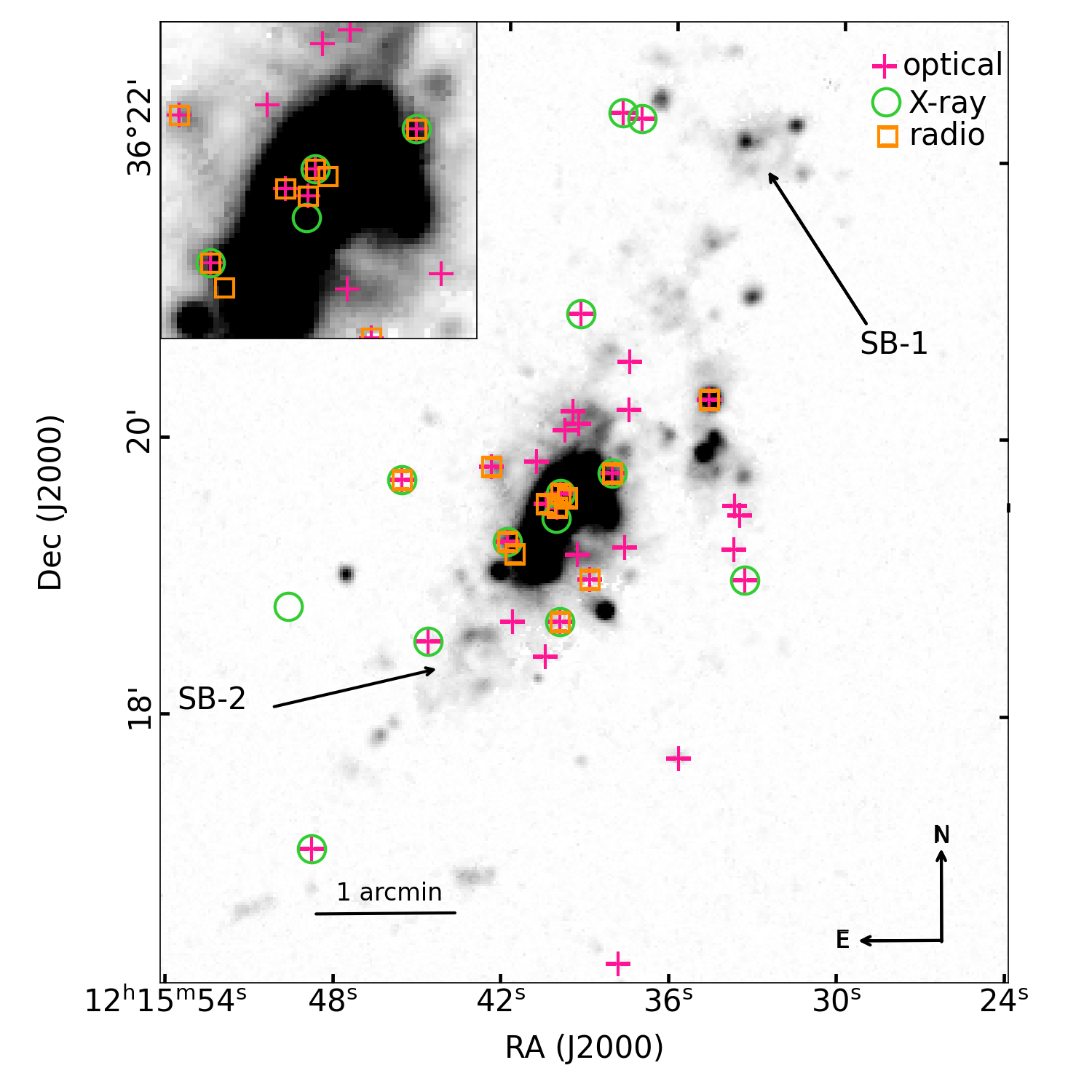}
    \caption{H$\alpha$ intensity map of NGC\,4214 with the position of its supernova remnants.  Pink crosses represent the SNRs with optical
emission, green circles represent the SNRs with X-ray emission, orange boxes are the radio SNRs.}
    \label{Fig10}
\end{figure}

Also, in Fig. \ref{Fig10} we show the H$\alpha$ monochromatic image of NGC\,4214 obtained from our FP cube with the 35 SNRs positions superimposed and pointed out with pink crosses the SNRs with optical emission, the green circles represent the SNRs with X-ray emission and with orange boxes the radio SNRs. From this figure we can see that the all SNRs are located in the bar region (see Section \ref{morpho}). Highlighting that seven SNRs (SNR-3, SNR-4, SNR-5, SNR-27, SNR-32 and SNR-34, SNR-35) are located inside the H II complex NGC\,4214-I.

With our FP observations we were able to study the kinematic of the SNRs in NGC\,4214 due to the spatial and spectral FP resolution. As we mentioned in the Introduction, the SNR sample used in this work was obtained from the SNRs detection in different wavelengths by previous works. Only the SNRs sample of \citet{Leonidaki2013} has information of their size; they consider that all the SNRs of their sample have a radius of about 1.4\arcsec (18.3 pc). Therefore, in this work we considered that all the SNRs of our sample have a radius of 18.3 pc. Then, we extracted the H$\alpha$ velocity profiles over windows of 3~pix $\times$ 3~pix (45.5\,pc $\times$ 45.5\,pc) and found that only 20 of the 35 SNRs has optimal S/N ratio.

\subsubsection{Expansion velocity}
\begin{figure*}
	\includegraphics[width=2\columnwidth]{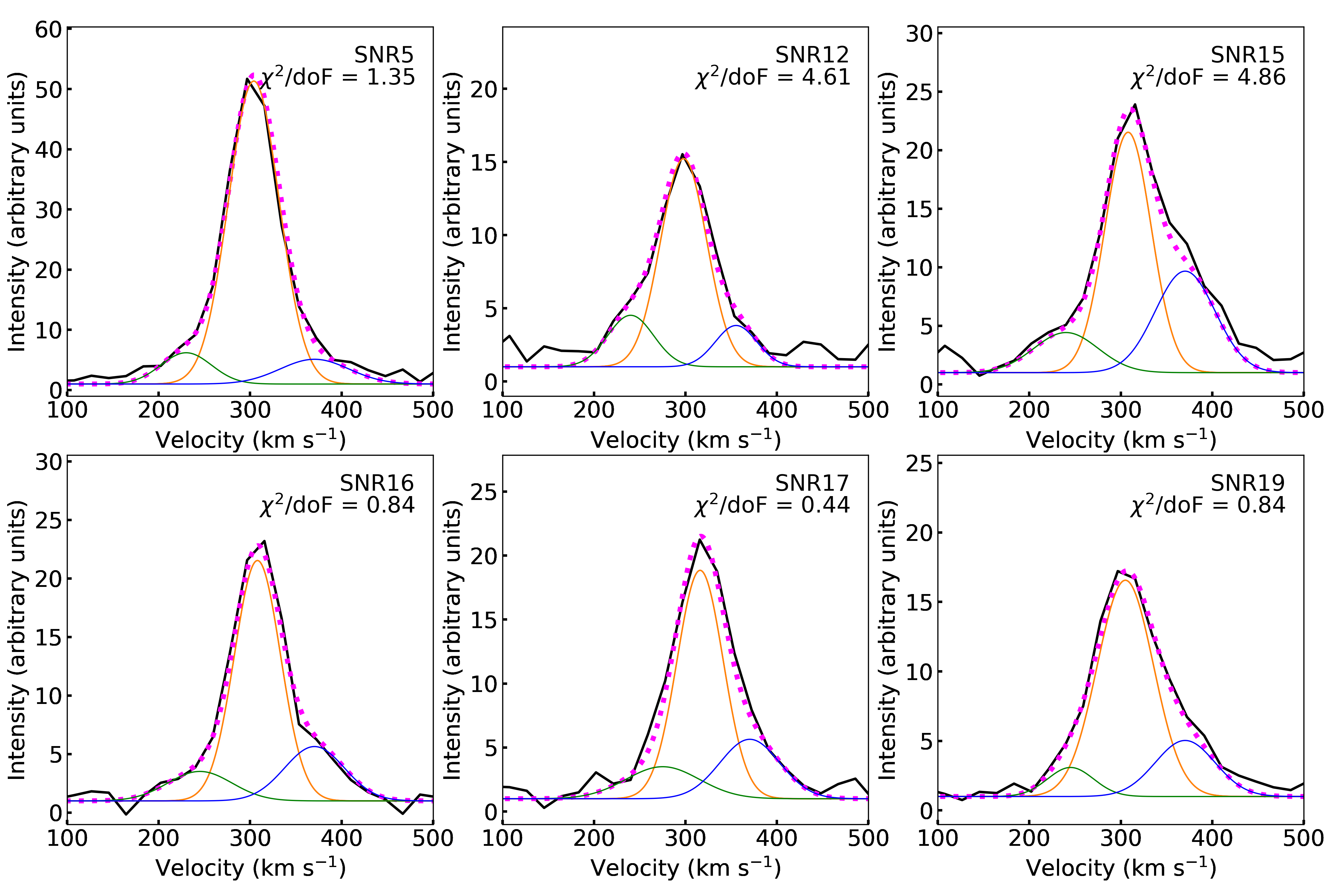}
    \caption{Examples of the fitted SNRs velocity profiles. In all the plots, the original profile is in black line. The orange line refers to the main velocity component, the secondary velocity components are in green and blue, and the best fit is the pink doted line.}
    \label{Fig11}
\end{figure*}

The expansion velocity of the SNRs was obtained fitting the velocity profiles of the SNRs with optimal S/N with the minimum of Gaussian functions.  All profiles present wings or humps as with SBs and similarity indicates the presence of composite profiles. Almost all SNRs profiles were fitted with three components except SNR-1 and SNR-23 which present multiple components and two components, respectively. For the velocity profiles fitted with three components, the main component (in orange) is associated with the rotation velocity of the gas around the centre of the galaxy. We show, as an example, the fitted velocity profile of six SNRs (SNR-5, SNR-12, SNR-15, SNR-16, SNR-17 and SNR-19) in Fig. \ref{Fig11}. The velocity components in green and blue are the blueshifted and redshift components, respectively, and are related to the expansion velocity of the SNR. Then, the expansion velocities were obtained considering the expansive  motion of a shell and assuming that the optical emission is not located at the centre of the SNR, therefore the expansion velocity is $V_\text{exp}$ = ($V_\text{max}$-$V_\text{min}$/2) where $V_\text{max}$ and $V_\text{min}$ are the blueshifted and redshifted components, respectively. 
For the SNR-1 and SNR-23, the expansion velocity was computed with their blueshifted and redshifted components. The expansion velocities of the SNRs with optimal S/N ratio are in general $V_\text{exp}$ > 100 km s$^{-1}$ except SNR-17 with $V_\text{exp}$ = 95 km s$^{-1}$. The list of SNRs and velocity values for each Gaussian component of the fitted velocity profiles, as well as their expansion velocities are shown in Table \ref{Table4}.

\begin{table}
	\centering
	\caption{Main parameters of SBs derived in this work.}
	\label{Table3}
	\begin{tabular}{lccr} 
		\hline
		Parameter & SB-1 & SB-2\\
		\hline
		size (pc$^2$) 			& 546.6$\times$ 440.2	& 516.2 $\times$ 409.9	\\
		Vexp (km s$^{-1}$)				& 47.5	& 47.5	\\
		n$_{0}$	(cm$^{-3}$)				& 0.58 $\pm$ 0.32 & 0.65$\pm$ 0.22 	\\
		E$_\text{mec}$ ($\times$10$^{51}$ erg)	   & 9.55 $\pm$ 5.40	& 8.68 $\pm$ 2.99 	\\
		t (10$^6$ yr)			& 2.22 $\pm$ 0.09	 & 2.07 $\pm$ 0.09 	\\
		L$_\text{W}$ (10$^{38}$ erg s$^{-1}$)	& 9.70 $\pm$ 5.42 & 9.47 $\pm$ 3.17	\\		
		M ($\times$10$^5$ M$_{\odot}$)					& 4.25$\pm$2.40	& 3.87 $\pm$ 1.33 	\\
		\hline
	\end{tabular}
\end{table}

\subsubsection{Electron density}
We determined the rms electron density (n$_\text{e}$) with the eq. \ref{eq1}, considering the linear radii for all SNRs was as 18.3 pc. The rms electron density values rank from 73 to 1000 cm$^{-3}$.

According to the SNR evolution diagram \citep[][]{Cioffi1990,Padmanabhan2001,Micelotta2018}, the SNRs are in the radiative phase \citep[see also][]{SanchezCruces2022} due to the expansion velocities we found (between 47 and 79 km s$^{-1}$) and their sizes (considered of about 18.3 pc). Therefore, to compute the pre-shock electron density n$_0$ we considered that the shock is radiative given by:

\begin{equation}
\label{eq4}
n_{0} = n_e\left(\frac{c_s}{V_s}\right)^2
\end{equation}

where c$_s$ = 10~km~s$^{-1}$ is the sound speed of the environment at T$_e$ = 10$^{4}$~K and V$_s$ = $V_{exp}$ is the shock velocity in km~s$^{-1}$. 
The pre-shock densities of SNRs rank from 18 to 257 cm$^{-3}$. We computed the uncertainty on $n_{0}$ using the error propagation to equation \ref{eq4}.

\subsubsection{The Energy and Age}
We can determine the age of the SNRs and the initial energy (E$_{0}$) deposited in the ISM by the SN explosion considering they are in the radiative phase with the  numerical model of \cite{Chevalier1974}:

\begin{equation}
\label{eq5}
t(4) =   30.7 R/V_s 
\end{equation}

\begin{equation}
\label{eq6}		
E_{0}(50) = 5.3\times10^{-7}n_0^{1.12}V_s^{1.4}R^{3.12}  
\end{equation}

where $V_\text{exp}$ is the shock velocity in km~s$^{-1}$, R is the linear radius in pc, t(4) is the age of the remnant in units of 10$^{4}$~yr and $E_\text{0}$(50) is in units of 10$^{50}$~erg. The linear radii for all SNRs was considered as 18.3 pc. The errors of t(4) and $E_{50}$ were computed applying the error propagation to equations \ref{eq5} and \ref{eq6}, respectively.  We were not able to obtain the age and initial energy of 15 SNRs due to poor detection S/N ratio (no expansion velocity information). 

The age of the SNRs found in this work, considering they are in the radiative phase, is about $\sim$10$^{4}$ yr and  the initial energy obtained is in the order of 10$^{50}$~erg, typical values of SNRs (see Table \ref{Table5}).

\section{Conclusions}

In this paper, we have presented H$\alpha$ high resolution Fabry-Perot observations of the galaxy NGC\,4214. This data was used to study the kinematics of the galaxy and locally study the kinematics of two new SBs and  SNRs in this galaxy.  Our main results and conclusions can be briefly summarized as follows:

We obtained the rotation curve of NGC\,4214 using the same kinematical parameters of \cite{Lelli2014} HI rotation curve. We found that in the innermost parts of the galaxy (R< 50 pc), the H$\alpha$ RC has an asymmetric behaviour indicating non-circular velocities in this region and that the kinematics of the ionised gas is probably dominated by the stellar bar motions. 

The H$\alpha$ and HI data RC are in agreement despite their difference data resolution; HI data has a resolution of 30 arcsec while the FP data resolution is 1.16 arcsec, and that this galaxy is very close to face-on and has a strong warp plus complex non-circular motions found in the HI RC of \cite{Lelli2014}. 

We found that the highest velocity dispersion are also related with the \ion{H}{II} complex with velocities of about $\sim$100 km~s$^{-1}$.

FP data allowed us for the first time to show the three-dimensional kinematic maps of the complexes NGC\,4214-I and NGC\,4214-II: i.e., we show  whole 2D fields of the complexes at different velocities.

The galaxy harbours two SBs detected in the H$\alpha$, [\ion{S}{ii}] $\lambda$6717 \AA, [\ion{N}{ii}] $\lambda$6584 \AA\ and [\ion{O}{iii}] $\lambda$5007 \AA\ direct images with sizes of 36\arcsec$\times$29\arcsec (546.6 pc$\times$440.3 pc) for SB-1 and 34\arcsec $\times$27\arcsec (516.2 pc $\times$ 409.9 pc) for SB-2. The expansion velocity of both SBs determined from their velocity profiles is 47.5 km~s$^{-1}$.  
 
The mechanical luminosity of the SBs  is L$_\text{W, SB-1}$ = 9.70$\pm$5.4$\times$10$^{38}$ erg s$^{-1}$ and for SB-2 L$_\text{W,SB-2}$ = 9.47$\pm$3.1$\times$10$^{38}$ erg s$^{-1}$ for SB-1 and SB-2, respectively. The dynamical ages of these SBs are about $\sim$2.0 Myr considering the energy conserving model indicating they are young SBs. 

We found that the SBs are probably formed by  massive stars or OB associations with stellar population similar to the N206 complex in the Large Magellanic Cloud with several WR stars plus more than 100 OB associations.  

Further studies of the stellar content of the SBs will help to determine whether or not WR stars and OB associations and understand the nature of these objects.

With our high spectral resolution observations, we were able to determine the expansion velocities and age of 20 of 35 SNRs hosted in this galaxy. The SNRs have a $V_\text{exp}$ between $\sim$48 to $\sim$80 km~s$^{-1}$, similar values obtained in 3C 400.2 ($V_\text{exp}$ = 60 km~s$^{-1}$) by \cite{Rosado1983a} and G206.9+2.3 ($V_\text{exp}$ = 86 km~s$^{-1}$) by  \cite{AmbrocioCruz2014}. The ages of the SNRs in NGC\,4214 are  between 7.1 and 11.8 $\times$10$^{4}$ yr, confirming they are SNRs. We also calculated the initial energy E$_{0}$ deposited in the ISM by the SN explosion of 20 of the 35 SNRs considering they are in the radiative phase. The initial energy ranges from 6.0 to 62.8 $\times$10$^{50}$ erg s$^{-1}$. Therefore, we confirmed that 20 of the 35 SNRs classified previously are SNRs.

\section*{Acknowledgements}
We would like to thank to the anonymous reviewer for the valuable comments and suggestions that helped in improving the article.
The authors acknowledge the financial support from Programa de Apoyo
a Proyectos de Investigación e Innovación Tecnológica (PAPIIT)
from the Universidad Nacional Autónoma de México (UNAM)
IN109919, Consejo Nacional de Ciencia y Tecnologia (CONACYT)
CY-253085 and CF-86367 grants. Based upon observations carried out
at the Observatorio Astronómico Nacional on the Sierra San Pedro
Mártir (OAN-SPM), Baja California, México. We thank the daytime
and night support staff at the OAN-SPM for facilitating and helping
obtain our observations. Facilities: OAN-SPM, México.

\section*{Data Availability}
The data underlying this article will be shared on reasonable request
to the corresponding author.



\bibliographystyle{mnras}
\bibliography{ngc4214} 





\bsp	
\label{lastpage}
\clearpage
\begin{landscape}
\begin{table}
\centering
\caption{Velocity information of the supernova remnants in NGC\,4214}
\label{Table4}
\begin{tabular}{lccccccccccccc} 
\hline
&&&
\multicolumn{9}{c}{Previous Identification} \\	
\cmidrule{4-12}
&&&
\multicolumn{3}{c}{Optical}&
\multicolumn{3}{c}{X-ray}&
\multicolumn{3}{c}{Radio}\\
&
\multicolumn{1}{c}{RA (J2000)}&
\multicolumn{1}{c}{Dec (J2000)}&
\multicolumn{3}{c}{\citep{Leonidaki2013}}&
\multicolumn{3}{c}{\citep{Leonidaki2010}}&
\multicolumn{3}{c}{\citep{Chomiuk2009}}\\
\cmidrule(lr){4-6}
\cmidrule(lr){7-9}
\cmidrule(lr){10-12}
\multicolumn{1}{c}{ID}&
\multicolumn{1}{c}{(h:m:s)}&
\multicolumn{1}{c}{($\degr$:$\arcmin$:$\arcsec$)}&
\multicolumn{1}{c}{ID} & 
\multicolumn{1}{c}{Clasiffication} &
\multicolumn{1}{c}{[\ion{S}{ii}]/H$\alpha$} & 
\multicolumn{1}{c}{ID} &
\multicolumn{1}{c}{Clasiffication} &
\multicolumn{1}{c}{L$_\text{X}^\text{unabs}$} & 
\multicolumn{1}{c}{ID} &
\multicolumn{1}{c}{Clasiffication} &
\multicolumn{1}{c}{$\alpha$}\\
\hline
SNR-1	&	12:15:38.96	&	+36:18:58.85	&	LBZ82	&	SNR[pc]	&	0.33$\pm$0.02	&	---	&	---	&	---	&	CW04	&	SNR	&	-0.48 $\pm$ 0.31	\\
SNR-2	&	12:15:40.00	&	+36:18:40.72	&	LBZ57	&	SNR[c]	&	0.92$\pm$0.09	&	LZB30	&	SNR	&	36.9	&	CW09	&	SNR	&	-0.62 $\pm$ 0.07	\\
SNR-3	&	12:15:40.01	&	+36:19:36.00	&	LBZ1098	&	SNR/\ion{H}{ii}	&	---	&	LZB34	&	SNR[p]	&	39.2	&	CW10	&	SNR	&	<-0.37	\\
SNR-4	&	12:15:40.14	&	+36:19:30.04	&	LBZ83	&	SNR[pc]	&	0.36$\pm$0.01	&	---	&	---	&	---	&	CW11	&	SNR	&	-0.53 $\pm$ 0.17	\\
SNR-5	&	12:15:41.90	&	+36:19:15.20	&	LBZ87	&	SNR[pc]	&	0.36$\pm$0.01	&	LZB28	&	SNR[p]	&	37.9	&	CW19	&	SNR	&	<-0.43	\\
SNR-6	&	12:15:33.60	&	+36:19:26.90	&	LBZ1	&	SNR	&	0.25$\pm$0.03	&	---	&	---	&	---	&	---	&	---	&	---	\\
SNR-7	&	12:15:33.80	&	+36:19:12.00	&	LBZ2	&	SNR	&	0.71$\pm$0.15	&	---	&	---	&	---	&	---	&	---	&	---	\\
SNR-8	&	12:15:33.80	&	+36:19:30.90	&	LBZ3	&	SNR	&	0.24$\pm$0.02	&	---	&	---	&	---	&	---	&	---	&	---	\\
SNR-9	&	12:15:35.70	&	+36:17:41.60	&	LBZ4	&	SNR	&	0.55$\pm$0.03	&	---	&	---	&	---	&	---	&	---	&	---	\\
SNR-10	&	12:15:37.60	&	+36:20:12.30	&	LBZ5	&	SNR	&	0.73$\pm$0.16	&	---	&	---	&	---	&	---	&	---	&	---	\\
SNR-11	&	12:15:37.60	&	+36:20:33.30	&	LBZ6	&	SNR	&	0.79$\pm$0.12	&	---	&	---	&	---	&	---	&	---	&	---	\\
SNR-12	&	12:15:37.70	&	+36:19:12.90	&	LBZ7	&	SNR	&	0.46$\pm$0.06	&	---	&	---	&	---	&	---	&	---	&	---	\\
SNR-13	&	12:15:37.80	&	+36:16:12.70	&	LBZ8	&	SNR	&	0.79$\pm$0.02	&	---	&	---	&	---	&	---	&	---	&	---	\\
SNR-14	&	12:15:39.40	&	+36:19:09.50	&	LBZ9	&	SNR	&	0.53$\pm$0.15	&	---	&	---	&	---	&	---	&	---	&	---	\\
SNR-15	&	12:15:39.40	&	+36:20:06.50	&	LBZ10	&	SNR	&	0.58$\pm$0.13	&	---	&	---	&	---	&	---	&	---	&	---	\\
SNR-16	&	12:15:39.60	&	+36:20:11.80	&	LBZ11	&	SNR	&	0.89$\pm$0.15	&	---	&	---	&	---	&	---	&	---	&	---	\\
SNR-17	&	12:15:39.90	&	+36:20:03.50	&	LBZ12	&	SNR	&	0.64$\pm$0.22	&	---	&	---	&	---	&	---	&	---	&	---	\\
SNR-18	&	12:15:40.50	&	+36:18:25.40	&	LBZ13	&	SNR	&	0.46$\pm$0.05	&	---	&	---	&	---	&	---	&	---	&	---	\\
SNR-19	&	12:15:40.90	&	+36:19:50.00	&	LBZ14	&	SNR	&	0.39$\pm$0.07	&	---	&	---	&	---	&	---	&	---	&	---	\\
SNR-20	&	12:15:41.70	&	+36:18:40.50	&	LBZ15	&	SNR	&	1.01$\pm$0.25	&	---	&	---	&	---	&	---	&	---	&	---	\\
SNR-21	&	12:15:42.50	&	+36:19:47.70	&	LBZ16	&	SNR	&	0.55$\pm$0.01	&	---	&	---	&	---	&	CW20	&	\ion{H}{ii}	&	0.10 $\pm$ 0.35	\\
SNR-22	&	12:15:44.70	&	+36:18:31.90	&	LBZ17	&	SNR	&	0.78$\pm$0.12	&	---	&	---	&	---	&	---	&	---	&	---	\\
SNR-23	&	12:15:45.70	&	+36:19:41.80	&	LBZ18	&	SNR	&	0.57$\pm$0.14	&	LZB38	&	SNR[p]	&	36.6	&	CW23	&	\ion{H}{ii}	&	0.10 $\pm$ 0.30	\\
SNR-24	&	12:15:33.41	&	+36:18:58.88	&	LBZ35*	&	SNR[c]	&	0.89$\pm$0.11	&	LZB7	&	SNR	&	36.9	&	---	&	---	&	---	\\
SNR-25	&	12:15:49.71	&	+36:18:46.69	&	---	&	---	&	---	&	LZB10	&	SNR[c]	&	--- 	&	---	&	---	&	---	\\
SNR-26	&	12:15:37.93	&	+36:22:21.00	&	LBZ47*	&	SNR[c]	&	0.63$\pm$0.02	&	LZB11	&	SNR[c]	&	--- 	&	---	&	---	&	---	\\
SNR-27	&	12:15:40.16	&	+36:19:25.21	&	---	&	---	&	---	&	LZB16	&	SNR[c]	&	--- 	&	---	&	---	&	---	\\
SNR-28	&	12:15:48.80	&	+36:17:01.83	&	LBZ73	&	SNR[c]	&	0.54$\pm$0.05	&	LZB23	&	SNR[c]	&	--- 	&	---	&	---	&	---	\\
SNR-29	&	12:15:39.37	&	+36:20:54.09	&	LBZ56	&	SNR[c]	&	0.89$\pm$0.21	&	LZB31	&	SNR[p]	&	36.6	&	---	&	---	&	---	\\
SNR-30	&	12:15:37.23	&	+36:22:18.65	&	LBZ936	&	SNR[c]	&	---	&	LZB35	&	SNR[p]	&	37.9	&	---	&	---	&	---	\\
SNR-31	&	12:15:34.74	&	+36:20:17.10	&	LBZ1073	&	SNR/\ion{H}{ii}	&	---	&	---	&	---	&	---	&	CW02	&	SNR	&	-0.48 $\pm$ 0.14	\\
SNR-32	&	12:15:40.55	&	+36:19:31.50	&	LBZ1099	&	LBZ1099	&	---	&	---	&	---	&	---	&	CW12	&	SNR	&	-0.56 $\pm$ 0.26	\\
SNR-33	&	12:15:38.18	&	+36:19:44.90	&	LBZ80	&	SNR[pc]	&	0.35$\pm$0.02	&	LZB26	&	XRB	&	38.0	&	CW03	&	SNR/\ion{H}{ii}	&	<0.00	\\
SNR-34	&	12:15:39.78	&	+36:19:34.30	&	---	&	---	&	---	&	---	&	---	&	---	&	CW08	&	SNR/\ion{H}{ii}	&	<-0.08	\\
SNR-35	&	12:15:41.64	&	+36:19:09.70	&	---	&	---	&	---	&	---	&	---	&	---	&	CW18	&	SNR/\ion{H}{ii}	&	<0.00	\\
\hline
\end{tabular}\\
\begin{flushleft}
Column 1 : SNR identification\\
Columns 2-3 : SNR right ascension and declination (epoch 2000).\\
Column 4-6 :  Identification, Classification and [\ion{S}{ii}]/H$\alpha$ given by \citet{Leonidaki2013}.\\
Columns 7-9 : Identification, Classification and the X-ray luminosity in erg s$^{-1}$ given by \cite{Leonidaki2010}. \\
Column 10-12 : Identification, Classification and the spectral indices (defined as S $\propto \nu^{\alpha}$) given by \cite{Chomiuk2009}. \\
Notation: SNR[c]- Candidate SNR. SNR[p]-Probable SNR, SNR[pc]-Probable candidate SNR, XRB-X-ray
binary.  SNR/\ion{H}{ii} in \citet{Leonidaki2013} are sources identified as SNRs in the X-ray or radio band but present ([\ion{S}{ii}]/H$\alpha$)$_\text{phot}$ < 0.3. SNR/\ion{H}{ii} in \cite{Chomiuk2009} are sources with H$\alpha$ associated and spectral index upper limit consistent with either an \ion{H}{ii} region or SNR.\\
\end{flushleft}	
\end{table}
\end{landscape}

\clearpage


\begin{table*}
\centering
\caption{Physical parameters of supernova remnants of NGC\,4214.}
\label{Table5}
\begin{tabular}{lccccccccc} 
\hline
ID &
\multicolumn{1}{c}{n$_e$ (rms)} &
\multicolumn{1}{c}{n$_0$} &
\multicolumn{1}{c}{V} &
\multicolumn{1}{c}{V$_\text{min}$} &
\multicolumn{1}{c}{V$_\text{max}$} &
\multicolumn{1}{c}{V$_\text{exp}$} &
\multicolumn{1}{c}{t(4)} &
\multicolumn{1}{c}{E$_{0}$(50)}\\
&
\multicolumn{1}{c}{(cm$^{-3}$)} &
\multicolumn{1}{c}{(cm$^{-3}$)} &
\multicolumn{1}{c}{(km s$^{-1}$)} &
\multicolumn{1}{c}{(km s$^{-1}$)} &
\multicolumn{1}{c}{(km s$^{-1}$)} &
\multicolumn{1}{c}{(km s$^{-1}$)} &
\multicolumn{1}{c}{(10$^{4}$ yr)} &
\multicolumn{1}{c}{(10$^{50}$ erg)} \\
\hline
SNR-1	&	181.37	&	3.44$\pm$0.19	&	---	&	235.14	&	380.40	&	72.63	&	7.74$\pm$0.21	&	7.41$\pm$0.17	\\
SNR-2	&	164.69	&	4.96$\pm$0.34	&	280.10	&	225.19	&	340.34	&	57.57	&	9.75$\pm$0.34	&	8.07$\pm$0.24	\\
SNR-3	&	715.44	&	12.7$\pm$0.70	&	300.05	&	230.12	&	380.30	&	75.09	&	7.48$\pm$0.20	&	33.5$\pm$0.70	\\
SNR-4	&	1028.14	&	19.5$\pm$1.10	&	304.14	&	235.11	&	380.40	&	72.64	&	7.74$\pm$0.21	&	51.7$\pm$1.20	\\
SNR-5	&	322.98	&	6.57$\pm$0.38	&	304.11	&	230.05	&	370.16	&	70.05	&	8.01$\pm$0.23	&	14.6$\pm$0.30	\\
SNR-9	&	162.14	&	2.59$\pm$0.13	&	256.55	&	165.25	&	323.36	&	79.05	&	7.10$\pm$0.18	&	6.08$\pm$0.13	\\
SNR-12	&	175.92	&	5.27$\pm$0.36	&	298.35	&	240.01	&	355.60	&	57.79	&	9.72$\pm$0.34	&	8.67$\pm$0.25	\\
SNR-14	&	245.27	&	5.79$\pm$0.36	&	308.15	&	235.24	&	365.39	&	65.07	&	8.63$\pm$0.27	&	11.4$\pm$0.30	\\
SNR-15	&	215.70	&	5.09$\pm$0.31	&	308.08	&	240.12	&	370.29	&	65.08	&	8.63$\pm$0.27	&	9.85$\pm$0.25	\\
SNR-16	&	165.95	&	4.25$\pm$0.27	&	308.25	&	245.11	&	370.12	&	62.50	&	8.99$\pm$0.29	&	7.60$\pm$0.20	\\
SNR-17	&	209.48	&	9.17$\pm$0.77	&	316.75	&	275.05	&	370.56	&	47.75	&	11.8$\pm$0.50	&	12.4$\pm$0.40	\\
SNR-19	&	206.64	&	5.24$\pm$0.33	&	305.20	&	245.09	&	370.77	&	62.84	&	8.95$\pm$0.28	&	9.68$\pm$0.26	\\
SNR-21	&	248.51	&	5.88$\pm$0.36	&	294.45	&	240.27	&	370.27	&	65.00	&	8.64$\pm$0.27	&	11.6$\pm$0.30	\\
SNR-23	&	191.44	&	3.40$\pm$0.18	&	---	&	300.00	&	450.00	&	75.00	&	7.49$\pm$0.20	&	7.66$\pm$0.17	\\
SNR-27	&	483.39	&	9.84$\pm$0.56	&	305.05	&	235.11	&	375.40	&	70.14	&	8.01$\pm$0.23	&	22.9$\pm$0.50	\\
SNR-31	&	751.00	&	12.5$\pm$0.60	&	275.05	&	200.11	&	355.40	&	77.64	&	7.24$\pm$0.19	&	34.4$\pm$0.70	\\
SNR-32	&	947.17	&	35.6$\pm$2.80	&	290.05	&	230.11	&	333.40	&	51.64	&	10.9$\pm$0.40	&	62.8$\pm$2.00	\\
SNR-33	&	416.13	&	8.47$\pm$0.48	&	305.05	&	230.11	&	370.40	&	70.14	&	8.01$\pm$0.23	&	19.3$\pm$0.50	\\
SNR-34	&	805.51	&	17.6$\pm$1.00	&	300.05	&	230.11	&	365.40	&	67.64	&	8.31$\pm$0.25	&	41.8$\pm$1.00	\\
SNR-35	&	275.32	&	5.60$\pm$0.32	&	300.05	&	230.11	&	370.40	&	70.14	&	8.01$\pm$0.23	&	12.2$\pm$0.30	\\
\hline
\end{tabular}\\
\begin{flushleft}
The columns are as follows:\\
Column 1 : SNR identification.\\
Column 2 : rms electron density obtained with eq. \ref{eq1}. \\
Column 3 : pre-shock electron density considering that the shock is radiative. \\
Column 4 : Main velocity component obtained from the fitted profiles.\\
Columns 5-6 : Blueshifted and redshifted components, (V$_\text{min}$ and V$_\text{max}$, respectively) from the fitted profiles.\\
Column 7 : Expansion velocity measured as $V_\text{exp}$ = (V$_\text{min}$ -V$_\text{max}$)/2. \\
Column 8 : Dynamical age computed using the numerical model of \citet{Chevalier1974}. \\
Column 9 : Initial energy deposited in the ISM by the SN explosion computed using the numerical model of \citet{Chevalier1974}.\\
\end{flushleft}
\end{table*}


\end{document}